%% file: main.tex
\pdfoutput=1

\documentclass[12pt,a4paper]{article}
\usepackage{ifthen} 
\newboolean{pdflatex}
\setboolean{pdflatex}{true} 

\newboolean{articletitles}
\setboolean{articletitles}{true} 

\newboolean{uprightparticles}
\setboolean{uprightparticles}{false} 

\newboolean{inbibliography}
\setboolean{inbibliography}{false} 

\input{preamble}
\usepackage{longtable} 

\begin{document}

\renewcommand{\thefootnote}{\fnsymbol{footnote}}
\setcounter{footnote}{1}

\input{title-LHCb-PAPER}




\pagestyle{plain}
\setcounter{page}{1}
\pagenumbering{arabic}


\input{introduction}

\input{angular}

\input{detector}

\input{selection}

\input{massmodel}

\input{fitmodel}

\input{results}

\input{acp}

\input{conclusions}

\input{acknowledgements}

\addcontentsline{toc}{section}{References}
\setboolean{inbibliography}{true}
\bibliographystyle{LHCb}
\bibliography{main,LHCb-PAPER,LHCb-CONF,LHCb-DP,local}

\end{document}

%% file: preamble.tex
\usepackage{microtype}
\usepackage{lineno}  
\usepackage{xspace} 
\usepackage{caption} 

\usepackage{graphicx}  
\usepackage{color}
\usepackage{colortbl}
\graphicspath{{./figs/}} 

\usepackage{amsmath} 
\usepackage{amssymb}
\usepackage{amsfonts}
\usepackage{upgreek} 

\newcommand*\patchAmsMathEnvironmentForLineno[1]{%
\expandafter\let\csname old#1\expandafter\endcsname\csname #1\endcsname
\expandafter\let\csname oldend#1\expandafter\endcsname\csname
end#1\endcsname
 \renewenvironment{#1}%
   {\linenomath\csname old#1\endcsname}%
   {\csname oldend#1\endcsname\endlinenomath}%
}
\newcommand*\patchBothAmsMathEnvironmentsForLineno[1]{%
  \patchAmsMathEnvironmentForLineno{#1}%
  \patchAmsMathEnvironmentForLineno{#1*}%
}
\AtBeginDocument{%
\patchBothAmsMathEnvironmentsForLineno{equation}%
\patchBothAmsMathEnvironmentsForLineno{align}%
\patchBothAmsMathEnvironmentsForLineno{flalign}%
\patchBothAmsMathEnvironmentsForLineno{alignat}%
\patchBothAmsMathEnvironmentsForLineno{gather}%
\patchBothAmsMathEnvironmentsForLineno{multline}%
}

\usepackage{hyperref}    
\usepackage[all]{hypcap} 

\input{lhcb-symbols-def} 

\input{extra-symbols}

\usepackage{cite} 
\usepackage{mciteplus}

%% file: lhcb-symbols-def.tex
\def\lhcb {\mbox{LHCb}\xspace}

\def\babar  {\mbox{BaBar}\xspace}
\def\belle  {\mbox{Belle}\xspace}

\ifthenelse{\boolean{uprightparticles}}%
{

 \def\Ppi         {\ensuremath{\uppi}\xspace}

 \def\PDelta      {\ensuremath{\Delta}\xspace}                 
 \def\PXi      {\ensuremath{\Xi}\xspace}                 
 \def\PLambda      {\ensuremath{\Lambda}\xspace}                 
 \def\PSigma      {\ensuremath{\Sigma}\xspace}                 
 \def\POmega      {\ensuremath{\Omega}\xspace}                 
 \def\PUpsilon      {\ensuremath{\Upsilon}\xspace}                 
 
 \def\PB      {\ensuremath{\mathrm{B}}\xspace}                 
                  
 \def\PD      {\ensuremath{\mathrm{D}}\xspace}

 \def\PK      {\ensuremath{\mathrm{K}}\xspace}

 \def\Pb      {\ensuremath{\mathrm{b}}\xspace}                 
 \def\Pc      {\ensuremath{\mathrm{c}}\xspace}

 \def\Pi      {\ensuremath{\mathrm{i}}\xspace}

 \def\Pp      {\ensuremath{\mathrm{p}}\xspace}

 \def\Ps      {\ensuremath{\mathrm{s}}\xspace}

}
{

 \def\Ppi         {\ensuremath{\pi}\xspace}

 \mathchardef\PDelta="7101
 \mathchardef\PXi="7104
 \mathchardef\PLambda="7103
 \mathchardef\PSigma="7106
 \mathchardef\POmega="710A
 \mathchardef\PUpsilon="7107
                  
 \def\PB      {\ensuremath{B}\xspace}                 
                  
 \def\PD      {\ensuremath{D}\xspace}

 \def\PK      {\ensuremath{K}\xspace}

 \def\Pb      {\ensuremath{b}\xspace}                 
 \def\Pc      {\ensuremath{c}\xspace}

 \def\Pi      {\ensuremath{i}\xspace}

 \def\Pp      {\ensuremath{p}\xspace}

 \def\Ps      {\ensuremath{s}\xspace}

}

\def\squark    {{\ensuremath{\Ps}}\xspace}

\def\cquark    {{\ensuremath{\Pc}}\xspace}

\def\bquark    {{\ensuremath{\Pb}}\xspace}

\def\pion   {{\ensuremath{\Ppi}}\xspace}

\def\kaon    {{\ensuremath{\PK}}\xspace}
  \def\Kbar    {{\kern 0.2em\overline{\kern -0.2em \PK}{}}\xspace}

\def\Kp      {{\ensuremath{\kaon^+}}\xspace}

\def\Kstarz  {{\ensuremath{\kaon^{*0}}}\xspace}
\def\Kstarzb {{\ensuremath{\Kbar^{*0}}}\xspace}

  \def\Dbar    {{\kern 0.2em\overline{\kern -0.2em \PD}{}}\xspace}

\def\B       {{\ensuremath{\PB}}\xspace}
\def\Bbar    {{\ensuremath{\kern 0.18em\overline{\kern -0.18em \PB}{}}}\xspace}

\def\Bz      {{\ensuremath{\B^0}}\xspace}
\def\Bzb     {{\ensuremath{\Bbar^0}}\xspace}

\def\Bd      {{\ensuremath{\B^0}}\xspace}
\def\Bs      {{\ensuremath{\B^0_\squark}}\xspace}
\def\Bsb     {{\ensuremath{\Bbar^0_\squark}}\xspace}
\def\Bdb     {{\ensuremath{\Bbar^0}}\xspace}

  \def\Y#1S{\ensuremath{\PUpsilon{(#1S)}}\xspace}

\def\proton      {{\ensuremath{\Pp}}\xspace}

\def\Lbar        {{\ensuremath{\kern 0.1em\overline{\kern -0.1em\PLambda}}}\xspace}

\newcommand{\decay}[2]{\ensuremath{#1\!\to #2}\xspace}         

\def\to                 {\ensuremath{\rightarrow}\xspace}

\def\CP                {{\ensuremath{C\!P}}\xspace}
\def\CPT               {{\ensuremath{C\!PT}}\xspace}

\def\AT#1     {\ensuremath{A_{\mathrm{T}}^{#1}}\xspace}           

\def\C#1      {\ensuremath{\mathcal{C}_{#1}}\xspace}                       
\def\Cp#1     {\ensuremath{\mathcal{C}_{#1}^{'}}\xspace}                    
\def\Ceff#1   {\ensuremath{\mathcal{C}_{#1}^{\mathrm{(eff)}}}\xspace}        
\def\Cpeff#1  {\ensuremath{\mathcal{C}_{#1}^{'\mathrm{(eff)}}}\xspace}       
\def\Ope#1    {\ensuremath{\mathcal{O}_{#1}}\xspace}                       
\def\Opep#1   {\ensuremath{\mathcal{O}_{#1}^{'}}\xspace}                    

\newcommand{\tev}{\ifthenelse{\boolean{inbibliography}}{\ensuremath{~T\kern -0.05em eV}\xspace}{\ensuremath{\mathrm{\,Te\kern -0.1em V}}}\xspace}
\newcommand{\gev}{\ensuremath{\mathrm{\,Ge\kern -0.1em V}}\xspace}
\newcommand{\mev}{\ensuremath{\mathrm{\,Me\kern -0.1em V}}\xspace}
\newcommand{\kev}{\ensuremath{\mathrm{\,ke\kern -0.1em V}}\xspace}
\newcommand{\ev}{\ensuremath{\mathrm{\,e\kern -0.1em V}}\xspace}
\newcommand{\gevc}{\ensuremath{{\mathrm{\,Ge\kern -0.1em V\!/}c}}\xspace}
\newcommand{\mevc}{\ensuremath{{\mathrm{\,Me\kern -0.1em V\!/}c}}\xspace}
\newcommand{\gevcc}{\ensuremath{{\mathrm{\,Ge\kern -0.1em V\!/}c^2}}\xspace}
\newcommand{\gevgevcccc}{\ensuremath{{\mathrm{\,Ge\kern -0.1em V^2\!/}c^4}}\xspace}
\newcommand{\mevcc}{\ensuremath{{\mathrm{\,Me\kern -0.1em V\!/}c^2}}\xspace}

\def\mum  {\ensuremath{{\,\upmu\rm m}}\xspace}

\def\invfb   {\ensuremath{\mbox{\,fb}^{-1}}\xspace}

\newcommand{\chisq}{\ensuremath{\chi^2}\xspace}

\newcommand{\chisqip}{\ensuremath{\chi^2_{\rm IP}}\xspace}

\def\gsim{{~\raise.15em\hbox{$>$}\kern-.85em
          \lower.35em\hbox{$\sim$}~}\xspace}
\def\lsim{{~\raise.15em\hbox{$<$}\kern-.85em
          \lower.35em\hbox{$\sim$}~}\xspace}

\newcommand{\Imag}{\ensuremath{\mathcal{I}m}\xspace}

\def\sPlot{\mbox{\em sPlot}}

\def\pt         {\mbox{$p_{\rm T}$}\xspace}

\def\dllkpi     {\ensuremath{\mathrm{DLL}_{\kaon\pion}}\xspace}

\def\evtgen     {\mbox{\textsc{EvtGen}}\xspace}

\def\geant      {\mbox{\textsc{Geant4}}\xspace}

\def\photos     {\mbox{\textsc{Photos}}\xspace}

\def\pythia     {\mbox{\textsc{Pythia}}\xspace}

\def\tell1  {TELL1\xspace}
\def\ukl1   {UKL1\xspace}



%% file: extra-symbols.tex
\def\BdPhiKst   {\decay{\Bd}{\phi \Kstarz}}
\def\BdPhiKstb  {\decay{\Bdb}{\phi \Kstarzb}}
\def\BsPhiKstb  {\decay{\Bsb}{\phi \Kstarz}}

\newcommand{\KKKpi}{\ensuremath{K^+K^-K^+\pi^-}\xspace}

\newcommand{\dper}{\delta_{\perp}\xspace}
\newcommand{\dpar}{\delta_{\parallel}\xspace}
\newcommand{\dsKpi}{\delta_{\textrm{S}}^{K\pi}\xspace}
\newcommand{\dsKK}{\delta_{\textrm{S}}^{KK}\xspace}

\newcommand{\Azero}{\ensuremath{A_{0}}\xspace}
\newcommand{\Aperp}{\ensuremath{A_{\perp}}\xspace}
\newcommand{\Apar}{\ensuremath{A_{\parallel}}\xspace}
\newcommand{\Azerob}{\ensuremath{\overline{A}_{0}}\xspace}
\newcommand{\Aperpb}{\ensuremath{\overline{A}_{\perp}}\xspace}
\newcommand{\Aparb}{\ensuremath{\overline{A}_{\parallel}}\xspace}
\newcommand{\AsKpi}{\ensuremath{A_{\textrm{S}}^{K\pi}}\xspace}
\newcommand{\AsKpib}{\ensuremath{\overline{A}_{\textrm{S}}^{K\pi}}\xspace}
\newcommand{\AsKK}{\ensuremath{A_{\textrm{S}}^{KK}}\xspace}
\newcommand{\AsKKb}{\ensuremath{\overline{A}_{\textrm{S}}^{KK}}\xspace}

%% file: title-LHCb-PAPER.tex
\begin{titlepage}
\pagenumbering{roman}

\vspace*{-1.5cm}
\centerline{\large EUROPEAN ORGANIZATION FOR NUCLEAR RESEARCH (CERN)}
\vspace*{2.0cm}
\hspace*{-0.5cm}
\begin{tabular*}{\linewidth}{lc@{\extracolsep{\fill}}r}
\ifthenelse{\boolean{pdflatex}}
{\vspace*{-2.7cm}\mbox{\!\!\!\includegraphics[width=.14\textwidth]{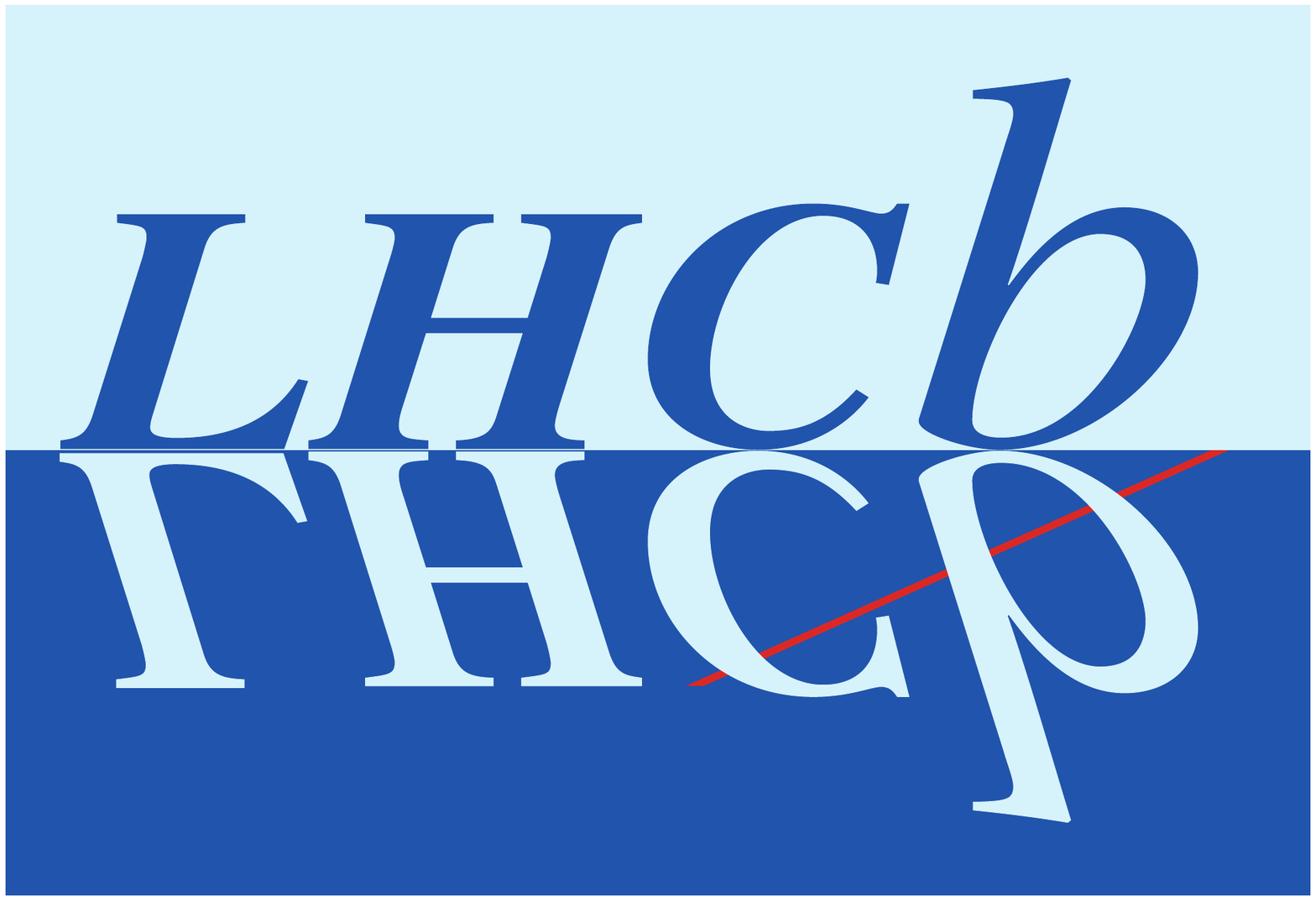}} & &}%
{\vspace*{-1.2cm}\mbox{\!\!\!\includegraphics[width=.12\textwidth]{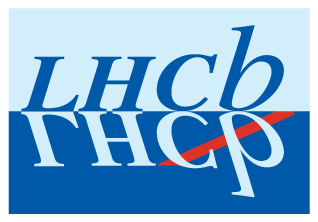}} & &}%
\\
 & & CERN-PH-EP-2014-038 \\
 & & LHCb-PAPER-2014-005 \\
 & & \today \\
 & & \\

\end{tabular*}

\vspace*{1.5cm}

{\bf\boldmath\huge
\begin{center}
 Measurement of polarization amplitudes and \CP
 asymmetries in $B^0 \to \phi K^*(892)^0$
\end{center}
}

\vspace*{1.0cm}

\begin{center}
The LHCb collaboration\footnote{Authors are listed on the following pages.}
\end{center}

\vspace{\fill}

\begin{abstract}
  \noindent An angular analysis of the decay $B^0 \to \phi K^*(892)^0$
  is reported based on a $pp$ collision data sample, corresponding to an integrated luminosity of $1.0\invfb$, collected at 
  a centre-of-mass energy of $\sqrt{s} = 7 \, \tev$ with the LHCb detector.
  The P-wave amplitudes and phases are measured with a greater 
  precision than by previous experiments, and confirm about equal amounts of longitudinal 
  and transverse polarization. The S-wave $K^+\pi^-$ and $K^+K^-$ contributions 
  are taken into account and found to be significant. A comparison of the $B^0 \to \phi K^*(892)^0$
  and $\kern 0.2em \overline{\kern -0.2em B}{}\xspace^0 \to \phi \kern 0.2em \overline{\kern -0.2em K}{}\xspace^*(892)^0$ results shows no evidence 
  for direct $C\!P$ violation in the rate asymmetry, in the triple-product 
  asymmetries or in the polarization amplitudes and phases.
\end{abstract}

\vspace*{2.0cm}

\begin{center}
  Submitted to the Journal of High Energy Physics
\end{center}

\vspace{\fill}

{\footnotesize 
\centerline{\copyright~CERN on behalf of the \lhcb collaboration, license \href{http://creativecommons.org/licenses/by/3.0/}{CC-BY-3.0}.}}
\vspace*{2mm}

\end{titlepage}

\newpage

\setcounter{page}{2}
\mbox{~}
\newpage
\input{LHCb_HD_authorlist_2014-01-25_2.tex}

\cleardoublepage

%% file: LHCb_HD_authorlist_2014-01-25_2.tex
\centerline{\large\bf LHCb collaboration}
\begin{flushleft}
\small
R.~Aaij$^{41}$, 
A.~Abba$^{21,u}$, 
B.~Adeva$^{37}$, 
M.~Adinolfi$^{46}$, 
A.~Affolder$^{52}$, 
Z.~Ajaltouni$^{5}$, 
J.~Albrecht$^{9}$, 
F.~Alessio$^{38}$, 
M.~Alexander$^{51}$, 
S.~Ali$^{41}$
% ...

{\footnotesize \it
% ...
$ ^{5}$Clermont Universit\'{e}, Universit\'{e} Blaise Pascal, CNRS/IN2P3, LPC, Clermont-Ferrand, France\\
% ...
$ ^{9}$Fakult\"{a}t Physik, Technische Universit\"{a}t Dortmund, Dortmund, Germany\\
% ...
$ ^{21}$Sezione INFN di Milano, Milano, Italy\\
% ...
$ ^{37}$Universidad de Santiago de Compostela, Santiago de Compostela, Spain\\
$ ^{38}$European Organization for Nuclear Research (CERN), Geneva, Switzerland\\
% ...
$ ^{41}$Nikhef National Institute for Subatomic Physics, Amsterdam, The Netherlands\\
% ...
$ ^{46}$H.H. Wills Physics Laboratory, University of Bristol, Bristol, United Kingdom\\
% ...
$ ^{51}$School of Physics and Astronomy, University of Glasgow, Glasgow, United Kingdom\\
$ ^{52}$Oliver Lodge Laboratory, University of Liverpool, Liverpool, United Kingdom\\
% ...
$ ^{u}$Politecnico di Milano, Milano, Italy\\
}
\end{flushleft}

%% file: introduction.tex
\section{Introduction}
\label{sec:Introduction}
\begin{figure}[ht]
\begin{center}
\includegraphics[scale=0.7]{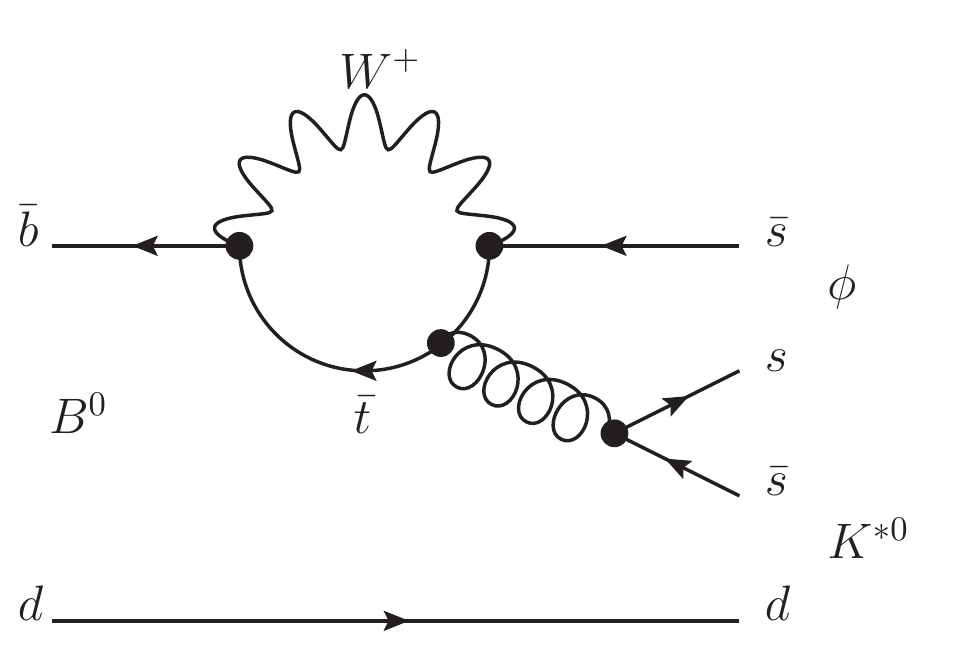}
\caption{\small Leading Feynman diagram for the \BdPhiKst decay.\label{fig:peng}}
\end{center}
\end{figure}

The decay \BdPhiKst\footnote{In this paper \Kstarz is defined as $K^*(892)^0$ unless otherwise stated.} has a branching fraction of $(9.8 \pm 0.6)\times
10^{-6}$~\cite{PDG2012}. In the Standard Model it proceeds mainly via
the gluonic penguin diagram shown in Fig.~\ref{fig:peng}. Studies of
observables related to \CP violation in this decay probe contributions from physics beyond the Standard Model in the 
penguin loop \cite{Datta:2003mj,PhysRevD.69.114013, gronau}. The decay was first observed by the CLEO
collaboration~\cite{Briere:2001ue}. Subsequently, branching fraction
measurements and angular analyses have been reported by the \babar and \belle collaborations~\cite{Aubert:2004xc, Aubert:2006uk, Aubert:2008zza, PhysRevLett.91.201801, PhysRevLett.94.221804, PhysRevD.88.072004}. 

The decay involves a spin-0 \B-meson decaying into two spin-1
vector mesons~($B \to VV$). Due to angular momentum conservation there are only three independent 
configurations of the final-state spin vectors, a longitudinal
component where in the $\Bz$ rest frame both resonances are polarized in their
direction of motion, and two transverse components with collinear and 
orthogonal polarizations. 
Angular analyses have shown that the longitudinal and transverse components in this decay
have roughly equal amplitudes. Similar results are seen in other $\B\rightarrow VV$ 
penguin transitions
\cite{delAmoSanchez:2010mz,Abe:2004mq, Aubert:2006fs,LHCb-PAPER-2011-012}. 
This is in contrast to tree-level decays such as
$\Bz\to\rho^+\rho^-$, where the $V-A$ nature of the weak interaction
causes the longitudinal component to dominate. 
The different behaviour of tree and penguin decays has attracted much
theoretical attention, with several explanations    
proposed such as large contributions from penguin annihilation effects \cite{kagan} or final-state
interactions \cite{PhysRevD.76.034015}. More recent
calculations based on QCD factorization \cite{beneke,PhysRevD.80.114026} are  
consistent with the data, although with significant uncertainties. 

In this paper, measurements of the polarization amplitudes,
phases, \CP asymmetries and triple-product asymmetries are presented. In the
Standard Model the \CP and triple-product asymmetries are expected to be small and were found to be consistent with zero by previous experiments~\cite{Aubert:2004xc, Aubert:2006uk, Aubert:2008zza, PhysRevLett.91.201801, PhysRevLett.94.221804}.  
The studies reported here are performed using $pp$ collision data, corresponding to an integrated 
luminosity of $1.0~\invfb$, collected at a centre-of-mass energy of 
$\sqrt{s} = 7 \, \tev$ with the \lhcb detector.

%% file: angular.tex
\section{Analysis strategy}
\label{sec:angular}
In this analysis the \BdPhiKst decay is studied, where the $\phi$
and \Kstarz mesons decay to $K^+ K^-$  and $K^+ \pi^-$, respectively
(the study of the charge conjugate \Bzb mode is implicitly assumed in this paper). 
Angular momentum conservation, for this  pseudoscalar to vector-vector transition, 
allows three possible helicity configurations of the vector-meson pair, 
with amplitudes denoted $H_{+1}$, $H_{-1}$ and $H_0$. These can be written as a longitudinal
polarization, $A_0$, and two transverse polarizations, $A_\perp$ and $A_\parallel$,
\begin{equation}
A_0  =  H_0\;, \hspace{1cm}
A_\perp = \frac{H_{+1} - H_{-1}}{\sqrt{2}}\;  \hspace{1cm} \mbox{and} \hspace{1cm}
A_\parallel = \frac{H_{+1} + H_{-1}}{\sqrt{2}}\;. 
\end{equation}
In addition to the dominant vector-vector~(P-wave) amplitudes, 
there are contributions where either the $K^+K^-$ or $K^+\pi^-$ pairs
are produced in a spin-0~(S-wave) state. These
amplitudes are denoted $A_{\textrm{S}}^{KK}$ and $A_{\textrm{S}}^{K\pi}$,
respectively. Only the relative phases of the amplitudes are physical observables. A phase convention is chosen such that $A_0$ is real. The remaining amplitudes have magnitudes and relative phases defined as
\begin{equation}
\Apar = |\Apar|e^{i\dpar}\;,  \hspace{0.1cm}  
\Aperp = |\Aperp|e^{i\dper}\;,  \hspace{0.1cm}
\AsKpi = |\AsKpi|e^{i\dsKpi} \hspace{0.1cm}  \mbox{and} \hspace{0.1cm} 
\AsKK = |\AsKK|e^{i\dsKK}\;.
\end{equation}
To determine these quantities, an analysis of
the angular distributions and invariant masses of the decay products
is performed. It is assumed that the contribution from $B^0 \to K^+K^-K^+\pi^-$, 
where both the $K^+K^-$ and $K^+\pi^-$ are non-resonant, is negligible.

In the following sections the key elements of the
analysis are discussed. First, the conventions used in the angular
analysis are defined together with the form of the differential
cross-section. Next, the parameterization of
the $K^+\pi^-$ and $K^+K^-$ mass distributions is discussed. Finally, the triple-product asymmetries that can be
derived from the angular variables are defined.

\subsection{Angular analysis}
\begin{figure}[htb]
\centering
\setlength{\unitlength}{1mm}
  \centering
  \begin{picture}(140,60)
    \put(0,-1){
      \includegraphics*[width=140mm,
      ]{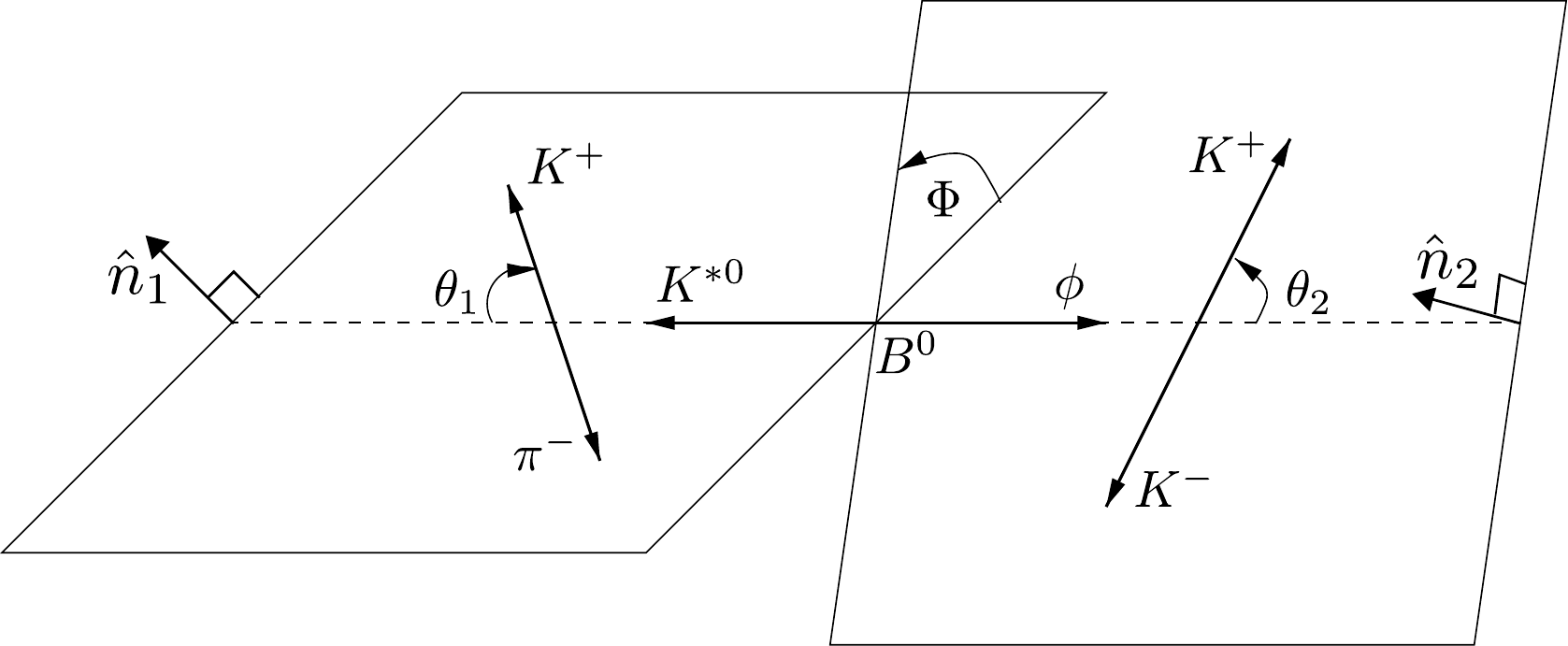}
    }
  \end{picture}
\caption{\label{fig:angles_def} The helicity angles $\theta_1$,
  $\theta_2$, $\Phi$ for the \BdPhiKst decay.}
\end{figure}

The angular analysis is performed in terms of three helicity angles
($\theta_1, \theta_2, \Phi$), as depicted in
Fig.~\ref{fig:angles_def}. The angle $\theta_1$ is defined as the
angle between the \Kp direction and the reverse of the \Bd direction in the \Kstarz rest frame. Similarly, $\theta_2$ is the
angle between the \Kp direction and the reverse of the \Bd direction in the $\phi$ rest frame. The angle $\Phi$ is the angle
between the decay planes of the $\phi$ and \Kstarz mesons in the \Bd rest frame.

The flavour of the decaying \Bd meson is determined by the
charge of the kaon from the \Kstarz decay. To determine the polarization amplitudes, the \Bz and \Bzb
decays are combined. 
For the study of \CP asymmetries, the \Bz and \Bzb decays are separated.

Taking into account both the P- and S-wave contributions and their
interference, the differential decay rate \cite{Aubert:2008zza} is given by the sum of
the fifteen terms given in Table~\ref{tab:decay_rate_terms},
\begin{equation}
d^5\Gamma
= \frac{9}{8\pi} \sum^{15}_{i=1} h_i~f_i( \theta_{1}, \theta_{2},
\Phi )  \mathcal{M}_i(m_{K\pi} ,m_{KK} )d\Omega(KKK\pi)\label{eqn:decay_sum}\;.
\end{equation}
The $h_i$ factors are combinations of the amplitudes, $f_i$ are functions of the helicity angles, $\mathcal{M}_i$ are functions of the
invariant mass of the intermediate resonances and $d\Omega(KKK\pi)$ is
a four-body phase-space factor, 
\begin{equation}
d\Omega(KKK\pi) \propto q_{\phi}q_{K^*}q_{\Bz}\hspace{0.5mm} d m_{K\pi}\hspace{0.5mm} d m_{KK}\hspace{0.5mm} d\textrm{cos}\hspace{0.5mm}{\theta_1}\hspace{0.5mm} d\textrm{cos}\hspace{0.5mm}{\theta_2}\hspace{0.5mm} d\Phi\;,
\end{equation}
where $q_A$ is the momentum of the daughter particles in the mother's
($A$ = $\Bz,\phi, \Kstarz$) centre-of-mass
system.
\begin{table}[htb]
\caption{\small Definition of the $h_i$, $f_i$ and $\mathcal{M}_i$ terms in Eq.~\ref{eqn:decay_sum}. Note that the P-wave interference terms $i=4$ and $i=6$ take the imaginary 
parts of $A_\perp A_\parallel^*$ and $A_\perp A_0^*$, while $i=5$ takes the real part of $A_\parallel A_0^*$. 
Similarly the S-wave interference terms $i=9$ and $i=13$ take the imaginary parts of $A_\perp A_\textrm{S}^* M_1 M_0^*$, 
and the terms $i=8,10,12,14$ take the real parts of $A_\parallel A_\textrm{S}^* M_1 M_0^*$ and $A_0 A_\textrm{S}^* M_1 M_0^*$.}\label{tab:decay_rate_terms}
\begin{center}
\resizebox{\columnwidth}{!}{%
\begin{tabular}{cccc}
\hline
$i$ & $h_i$                 & $f_i(\theta_1, \theta_2, \Phi)$                           & $\mathcal{M}_i(m_{K\pi},m_{KK})$ \\ \hline 
1 & $|A_0|^2$         & $\cos\theta_1^2 \cos\theta_2^2$                          & $|M_1^{K\pi}(m_{K\pi})|^2|M_1^{KK}(m_{KK})|^2$ \\
2 & $|A_\parallel|^2$ & $\frac{1}{4}\sin\theta_1^2\sin\theta_2^2(1+\cos(2\Phi))$ & $|M_1^{K\pi}(m_{K\pi})|^2|M_1^{KK}(m_{KK})|^2$ \\
3 & $|A_\perp|^2$     & $\frac{1}{4}\sin\theta_1^2\sin\theta_2^2(1-\cos(2\Phi))$ & $|M_1^{K\pi}(m_{K\pi})|^2|M_1^{KK}(m_{KK})|^2$ \\
4 & $|A_\perp| |A_\parallel^*|e^{i(\delta_\perp-\delta_\parallel)}$    & $-\frac{1}{2}\sin\theta_1^2\sin\theta_2^2\sin(2\Phi)$               & $|M_1^{K\pi}(m_{K\pi})|^2|M_1^{KK}(m_{KK})|^2$ \\
5 & $|A_\parallel| |A_0^*|e^{i\delta_\parallel}$        & $\sqrt{2}\cos\theta_1\sin\theta_1\cos\theta_2\sin\theta_2\cos\Phi$  & $|M_1^{K\pi}(m_{K\pi})|^2|M_1^{KK}(m_{KK})|^2$ \\
6 & $|A_\perp| |A_0^*|e^{i\delta_\perp}$            & $-\sqrt{2}\cos\theta_1\sin\theta_1\cos\theta_2\sin\theta_2\sin\Phi$ & $|M_1^{K\pi}(m_{K\pi})|^2|M_1^{KK}(m_{KK})|^2$ \\
7 & $|A_\textrm{S}^{K\pi}|^2$           & $\frac{1}{3} \cos\theta_2^2$                                        & $|M_0^{K\pi}(m_{K\pi})|^2|M_1^{KK}(m_{KK})|^2$ \\
8 & $|A_\parallel| |A_\textrm{S}^{*K\pi}|e^{i(\delta_\parallel-\delta_\textrm{S}^{K\pi})}$  & $\frac{\sqrt{6}}{3} \sin\theta_1\cos\theta_2\sin\theta_2\cos\Phi$   & $|M_1^{KK}(m_{KK})|^2 M_1^{K\pi}(m_{K\pi})M_0^{*K\pi}(m_{K\pi})$ \\
9 & $|A_\perp| |A_\textrm{S}^{*K\pi}|e^{i(\delta_\perp-\delta_\textrm{S}^{K\pi})}$       & $-\frac{\sqrt{6}}{3} \sin\theta_1\cos\theta_2\sin\theta_2\sin\Phi$  & $|M_1^{KK}(m_{KK})|^2 M_1^{K\pi}(m_{K\pi})M_0^{*K\pi}(m_{K\pi})$ \\
10 & $|A_0| |A_\textrm{S}^{*K\pi}|e^{-i\delta_\textrm{S}^{K\pi}}$         & $\frac{2}{\sqrt{3}}\cos\theta_1\cos\theta_2^2$                      & $|M_1^{KK}(m_{KK})|^2 M_1^{K\pi}(m_{K\pi})M_0^{*K\pi}(m_{K\pi})$ \\
11 & $|A_\textrm{S}^{KK}|^2$            & $\frac{1}{3} \cos\theta_1^2$                                        & $|M_0^{KK}(m_{KK})|^2|M_1^{K\pi}(m_{K\pi})|^2$ \\
12 & $|A_\parallel| |A_\textrm{S}^{*KK}|e^{i(\delta_\parallel-\delta_\textrm{S}^{KK})}$   & $\frac{\sqrt{6}}{3} \sin\theta_1\cos\theta_1\sin\theta_2\cos\Phi$   & $|M_1^{K\pi}(m_{K\pi})|^2 M_1^{KK}(m_{KK})M_0^{*KK}(m_{KK})$ \\
13 & $|A_\perp| |A_\textrm{S}^{*KK}|e^{i(\delta_\perp-\delta_\textrm{S}^{KK})}$       & $-\frac{\sqrt{6}}{3} \sin\theta_1\cos\theta_1\sin\theta_2\sin\Phi$  & $|M_1^{K\pi}(m_{K\pi})|^2 M_1^{KK}(m_{KK})M_0^{*KK}(m_{KK})$ \\
14 & $|A_0| |A_\textrm{S}^{*KK}|e^{-i\delta_\textrm{S}^{KK}}$           & $\frac{2}{\sqrt{3}}\cos\theta_1^2\cos\theta_2$                      & $|M_1^{K\pi}(m_{K\pi})|^2 M_1^{KK}(m_{KK})M_0^{*KK}(m_{KK})$ \\
15 & $|A_\textrm{S}^{K\pi}| |A_\textrm{S}^{*KK}|e^{i(\delta_\textrm{S}^{K\pi}-\delta_\textrm{S}^{KK})}$    & $\frac{2}{3}\cos\theta_1\cos\theta_2$ & $M_1^{KK}(m_{KK})M_0^{K\pi}(m_{K\pi})M_0^{*KK}(m_{KK}) M_1^{*K\pi}(m_{K\pi})$ \\ \hline
\end{tabular}
}
\end{center}
\end{table}

The differential decay rate for \BdPhiKstb is obtained by defining the angles using the
charge conjugate final state particles and multiplying the
interference terms $f_4, f_6, f_9, f_{13}$ by $-1$. To allow for direct \CP violation, the
amplitudes $A_j$ are replaced by $\overline{A}_j$, for
$j={0, \parallel, \perp,\textrm{S}}$. The rate is normalized separately for the \Bzb and
\Bz decays such that the P- and S-wave fractions are
\begin{equation}
F_\textrm{P} = |\Azero|^2 + |\Apar|^2 + |\Aperp|^2\;, \hspace{1cm}
F_\textrm{S} = |\AsKpi|^2 + |\AsKK|^2\;, \hspace{1cm}
F_\textrm{P} + F_\textrm{S} = 1\;,
\end{equation}
and
\begin{equation}
\overline{F}_\textrm{P} = |{\Azerob}|^2 + |{\Aparb}|^2 + |{\Aperpb}|^2\;, \hspace{1cm}
\overline{F}_\textrm{S} = |{\AsKpib}|^2 + |{\AsKKb}|^2\;, \hspace{1cm}
\overline{F}_\textrm{P} + \overline{F}_\textrm{S} = 1\;.  
\end{equation}
In addition, a convention is adopted such that the phases $\dsKpi$ and
$\dsKK$ are defined as the difference between the P- and S-wave phases
at the \Kstarz and $\phi$ meson
poles, respectively.
\subsection{Mass distributions}
\label{sec:massResonance}
The differential decay width depends on the invariant masses of
the $K^+\pi^-$ and $K^+K^-$ systems, denoted $m_{K\pi}$ and $m_{KK}$, respectively. The P-wave $K^+\pi^-$ amplitude is parameterized using a
relativistic spin-$1$ Breit-Wigner resonance function,
\begin{equation}
  M^{K\pi}_1(m_{K\pi}) =  \frac{m_{K\pi}}{q_{K^*}} \hspace{1mm} \frac{m_0^{K^*} \Gamma_1^{K\pi}(m_{K\pi})}{(m_0^{K^*})^2 -
m_{K\pi}^2 - im_0^{K^*}\Gamma_1^{K\pi}(m_{K\pi}) }\;,
\label{eq:M1propagator}
\end{equation}
where $m_0^{K^*} = 895.81 \, \mevcc$ \cite{PDG2012} is the
\Kstarz mass.
The mass-dependent width is given by
\begin{equation}
 \Gamma_1^{K\pi}(m_{K\pi}) = \Gamma_0^{K^*} \frac{m_0^{K^*}}{m_{K\pi}} \hspace{1mm}
 \frac{1+r^2 q_0^2}{1 + r^2 q_{K^*}^2} \left( \frac{q_{K^*}}{q_0}\right)^3\;,
\end{equation}
where $q_0$ is the value of $q_{K^*}$ at $m_0^{K^*}$, $r = 3.4~\hbar c/$GeV~\cite{LASS}
 is the interaction radius and $\Gamma_0^{K^*} = 47.4 \mevcc$ is
the natural width of the \Kstarz meson~\cite{PDG2012}. The P-wave $K^+K^-$ amplitude, denoted $M_1^{KK}(m_{KK})$, is modelled
in a similar way using the values $m_0^{\phi} = 1019.455 \,
\mevcc$ and $\Gamma_0^{\phi} = 4.26 \, \mevcc$~\cite{PDG2012}. In
the case of the $\phi$ meson the natural width is comparable to the
detector resolution of $1.2 \,\mevcc$, which is accounted for by convolving the
Breit-Wigner with a Gaussian function.

As the \Kstarz is a relatively broad resonance, the S-wave
component in the $K^+\pi^-$ system, denoted $M_0^{K\pi}(m_{K\pi})$,  needs careful treatment. In this
analysis the approach described in Ref.~\cite{Aubert:2008zza} is followed,
which makes use of the LASS parameterization \cite{LASS}. This takes
into account an $L=0$ $K^{*}_0(1430)$ contribution together with a
non-resonant amplitude. The values used for the LASS parameterization are taken from Ref.~\cite{Aubert:2008zza}.

Finally, an S-wave in the $K^+K^-$ system is considered. This is
described by the Flatt\'e parameterization of the $f_0(980)$ resonance\cite{flatte}, 
\begin{equation}
  M_0^{KK}(m_{KK}) = \frac{1}{m_{f_0}^2-m_{KK}^2 -i m_{f_0} ( g_{\pi\pi}
\rho_{\pi\pi}+ g_{KK} \rho_{KK} )}\;,
\label{formula:flatte}
\end{equation}
where the $g_{KK,\pi\pi}$ are partial decay widths and the $\rho_{KK,\pi\pi}$ are phase-space
factors. The values $m_{f_0} = 939\mevcc$, $g_{\pi\pi} = 199 \mevcc$ and
$g_{KK}/g_{\pi\pi} = 3.0$ were measured in Ref.~\cite{LHCb-PAPER-2012-045}. The Flatt\'e distribution is
convolved with a Gaussian function to account for the detector resolution. Other approaches 
to modelling the mass distributions for both the $K^+\pi^-$ and $K^+K^-$ S-wave are considered as part of the systematic uncertainty determination.
\subsection{Triple-product asymmetries}
The amplitudes and phases can be used to calculate triple-product
asymmetries~\cite{gronau, Bensalem:2000hq, Datta:2003mj}. Non-zero
triple-product asymmetries arise either due to a $T$-violating phase or a \CP-conserving phase
and final-state interactions. Assuming \CPT symmetry, a $T$-violating phase,
which is a \textit{true} asymmetry, implies that \CP is violated.  

For the P-wave decay, two triple-product asymmetries are calculated from the results of the angular
analysis~{\cite{gronau}},
\begin{equation}
A_{T}^{1} = \frac{\Gamma(s_{\theta_1\theta_2}\sin\Phi>0) - \Gamma(s_{\theta_1\theta_2}\sin\Phi<0)}{\Gamma(s_{\theta_1\theta_2}\sin\Phi>0) +
\Gamma(s_{\theta_1\theta_2}\sin\Phi<0)} 
\hspace{0.4cm} \mbox{and} \hspace{0.4cm}
A_{T}^{2} = \frac{\Gamma(\sin2\Phi>0) - \Gamma(\sin2\Phi<0)}{\Gamma(\sin2\Phi>0) + \Gamma(\sin2\Phi<0)}\;,
\end{equation}
where $s_{\theta_1\theta_2}=\textrm{sign}(\cos\theta_1\cos\theta_2)$. These asymmetries can be rewritten in terms of the interference terms 
between the amplitudes~\cite{gronau}, $h_4$ and $h_6$ in Table~\ref{tab:decay_rate_terms},
\begin{equation}
A_T^{1} = -\frac{4}{\pi}\Imag(A_{\perp}A_{0}^*)  \hspace{1cm}  \mbox{and} \hspace{1cm} 
A_T^{2} = -\frac{2\sqrt{2}}{\pi}\Imag(A_{\perp}A_{\parallel}^*)\;.
\label{for:TP12}
\end{equation}
Since the decay products identify the flavour at decay, the data can be
separated into $\Bz$ and $\Bzb$ decays and the triple-product
asymmetries calculated for both cases. This allows a determination of
the \textit{true} asymmetries, $A_{T}^{\, k}{\rm (true)} = ({A^k_T +
  \overline{A}^{\, k}_T})/2$,  and so called \textit{fake} asymmetries,
$A_{T}^k{\rm (fake)} = (A^k_T - \overline{A}^{\, k}_T)/2$, where $k = 1,2$. 
In the Standard Model
the value of $A_{T}^k$(true) is predicted to be zero and any
deviation from this would indicate physics beyond the Standard Model. Non-zero values for $A_{T}^k$(fake) 
reflect the importance of strong final-state phases \cite{gronau}. 

The S-wave contributions allow two additional triple-product asymmetries to be defined from $h_9$ and $h_{13}$ in Table~\ref{tab:decay_rate_terms},
\begin{align}
A_T^3 &= \frac{\displaystyle \Gamma(s_{\theta_1}\sin\Phi>0) - \Gamma(s_{\theta_1}\sin\Phi<0)}{\Gamma(s_{\theta_1}\sin\Phi>0) +
\Gamma(s_{\theta_1}\sin\Phi<0)} \nonumber \\ 
&= -\sqrt{\frac{3}{2}}\int |M_1^{KK}(m_{KK})|^2 \Imag(A_\perp
A_\textrm{S}^{*K\pi}M_1^{K\pi}(m_{K\pi})M_0^{*K\pi}(m_{K\pi}))dm_{KK}dm_{K\pi}\;,
\end{align}
and
\begin{align}
A_T^4 &= \frac{ \Gamma(s_{\theta_2}\sin\Phi>0) - \Gamma(s_{\theta_2}\sin\Phi<0)}{ \Gamma(s_{\theta_2}\sin\Phi>0) +
\Gamma(s_{\theta_2}\sin\Phi<0)}  \nonumber \\
&= -\sqrt{\frac{3}{2}}\int |M_1^{K\pi}(m_{K\pi})|^2 \Imag(A_\perp A_\textrm{S}^{*KK}M_1^{KK}(m_{KK})M_0^{*KK}(m_{KK}))dm_{KK}dm_{K\pi}\;,
\label{for:TP34}
\end{align}
where $s_{\theta_i}=\textrm{sign}(\cos\theta_i)$ for $i=1,2$.

%% file: detector.tex
\section{Detector and dataset}
\label{sec:Detector}
The \lhcb detector~\cite{Alves:2008zz} is a single-arm forward
spectrometer covering the \mbox{pseudorapidity} range $2<\eta <5$,
designed for the study of particles containing \bquark or \cquark
quarks. The detector includes a high-precision tracking system
consisting of a silicon-strip vertex detector surrounding the $pp$
interaction region, a large-area silicon-strip detector located
upstream of a dipole magnet with a bending power of about
$4{\rm\,Tm}$, and three stations of silicon-strip detectors and straw
drift tubes placed downstream. The 
polarity of the dipole magnet is reversed at intervals corresponding
to roughly $0.1 \invfb$ of 
collected data, in order to minimize systematic uncertainties associated with detector asymmetries. 
The combined tracking system provides a momentum measurement with
relative uncertainty that varies from 0.4\,\% at 5\gevc to 0.6\,\% at 100\gevc,
and impact parameter resolution of 20\mum for
tracks with high transverse momentum~(\pt). Charged hadrons are identified
using two ring-imaging Cherenkov
detectors~\cite{LHCb-DP-2012-003}. Photon, electron 
and hadron candidates are identified by a calorimeter system consisting of
scintillating-pad and preshower detectors, an electromagnetic
calorimeter and a hadronic calorimeter. Muons are identified by a
system composed of alternating layers of iron and multiwire
proportional chambers.

The trigger~\cite{LHCb-DP-2012-004} consists of a
hardware stage, based on information from the calorimeter and muon
systems, followed by a software stage, which applies a full event
reconstruction. In this analysis two categories of events that pass
the hardware trigger stage are considered: those where the signal
\bquark-hadron products are used in the trigger decision (TOS) and those
where the trigger decision is caused by other activity in the event
(TIS) \cite{LHCb-DP-2012-004}. The software trigger requires a three-track secondary vertex 
with large transverse momenta of the tracks and a 
significant displacement from the primary $pp$ interaction vertices~(PVs). 
At least one track should
have $\pt >1.7\gevc$ and \chisqip with respect to any 
primary interaction greater than 16, where \chisqip is defined as the 
difference in \chisq of a given PV reconstructed with and 
without the considered track. A multivariate algorithm~\cite{BBDT} is
used for the identification of secondary vertices consistent with the
decay of a \bquark hadron.

Simulated data samples are used to correct for the 
detector acceptance and response. In the simulation, $pp$ collisions are generated using
\pythia~6.4~\cite{Sjostrand:2006za} with a specific \lhcb
configuration~\cite{LHCb-PROC-2010-056}.  Decays of hadronic particles
are described by \evtgen~\cite{Lange:2001uf}, in which final-state
radiation is generated using \photos~\cite{Golonka:2005pn}. The
interaction of the generated particles with the detector and its
response are implemented using the \geant
toolkit~\cite{Allison:2006ve, *Agostinelli:2002hh} as described in
Ref.~\cite{LHCb-PROC-2011-006}.

%% file: selection.tex
\section{Event selection}
\label{sec:selection} 
The selection of events is divided into two
parts. In the first step a loose selection is performed that 
retains the majority of signal events, whilst reducing the background by
a large fraction. Following this, a multivariate method is used to further reduce the background. 

The selection starts from well reconstructed charged particles with a $\pt > 500 \, \mevc$
that traverse the entire spectrometer. Fake tracks, not associated to actual charged particles, are suppressed using the output 
of a neural network trained to discriminate between these and real particles~\cite{LHCb-PAPER-2013-010}. Further background suppression is achieved
by exploiting the fact that the products of \bquark-hadron decays have
a large impact parameter (IP) with respect to the nearest PV. The IP of each track with respect to any
primary vertex is required to have a $\chisqip>9$.

To select well-identified pions and kaons, the difference in the logarithms of the 
likelihood of the kaon hypothesis relative to the pion
hypothesis ($\dllkpi$) is provided using information from the
ring-imaging Cherenkov detectors. The kaons that form the $\phi\to
K^+K^-$ candidate are required to have $\dllkpi>0$. To reduce background from $\pi^+\pi^-$ pairs, a tighter requirement, $\dllkpi > 2$, is applied to the kaon in the $K^+\pi^-$ pair. For the pion in the $K^+\pi^-$ pair the requirement is $\dllkpi<0$.  

The resulting charged particles are combined to form $\phi$
and \Kstarz meson candidates.
The invariant mass of the $K^+K^-$
($K^+\pi^-) $ pair is required to be within $\pm 15 \mevcc$ ($\pm 150
\mevcc$) of the known mass of the $\phi$ (\Kstarz) meson \cite{PDG2012}. 
Finally, the $\pt$ of the $\phi$ and \Kstarz mesons should both be greater than $900 \mevc$, 
and the fit of their two-track vertices should have a $\chi^2<9$. 

Candidate \Bz meson decays with $\KKKpi$ invariant mass in the range
$5150<m_{KKK\pi}<5600\mevcc$ are formed from pairs of selected $\phi$
and \Kstarz meson candidates. 
A fit is made requiring all four final-state particles to originate 
from a common vertex and the $\chi^2$ per degree of freedom of this
fit is required to be less than 15. 
To remove $\Bs \rightarrow \phi \phi$ decays where a kaon has
been incorrectly identified as a pion, the invariant mass of the $K^+ \pi^-$ pair is
recalculated assuming that both particles are kaons. If the resulting invariant
mass is within $\pm 15 \mevcc$ of the known $\phi$ mass,
the candidate is rejected. Finally, the decay vertex of the \Bz meson candidate is required to be
displaced from the nearest PV, with a flight distance significance of more than 5 standard deviations, 
and the \Bz momentum vector is required to point back towards the PV with an impact parameter less than 
$0.3$~mm and $\chisqip<5$. 

Further background suppression is achieved using a geometric likelihood (GL) method~\cite{Karlen,LHCb-PAPER-2011-012,LHCb-PAPER-2013-012}. The GL is 
trained using a sample of simulated \BdPhiKst signal events together with background events selected from
the upper mass sideband of the $\B^0$ meson, $m_{KKK\pi}>5413\mevcc$, and the $\phi$ mass sidebands, 
$|m_{KK}-m_0^{\phi}|>15\mevcc$.
These sidebands are not used in the subsequent analysis. 
Six discriminating variables are input to the GL: the IP of the \Bz candidate with respect to the PV, the distance of closest
approach of the $\phi$ and $K^{*0}$ meson candidate trajectories, the lifetime of the
\Bz candidate, the transverse momentum of the \Bz candidate, the minimum $\chisqip$ of the $K^+K^-$ pair and the minimum $\chisqip$ of the $K^+ \pi^-$ pair. 
As a figure of merit the ratio
$S/\sqrt{S+B}$ is considered, where $S$ and $B$ are the yields of signal and background events
in the training samples, scaled to match the observed signal and background yields in the data. 
The maximum value for the figure of merit is found to be $24.6$ for $\textrm{GL}>0.1$, with signal and background efficiencies 
of $90 \, \%$ and $21 \,\%$, respectively, compared to the selection
performed without the GL. This reduces the sample size for the final analysis to 1852 candidates.

%% file: massmodel.tex
\section{\boldmath ${K^+K^-K^+\pi^-}$ mass model}
\label{sec:massmodel}
\begin{figure}[htb]
\begin{center}
\includegraphics[height=8cm]{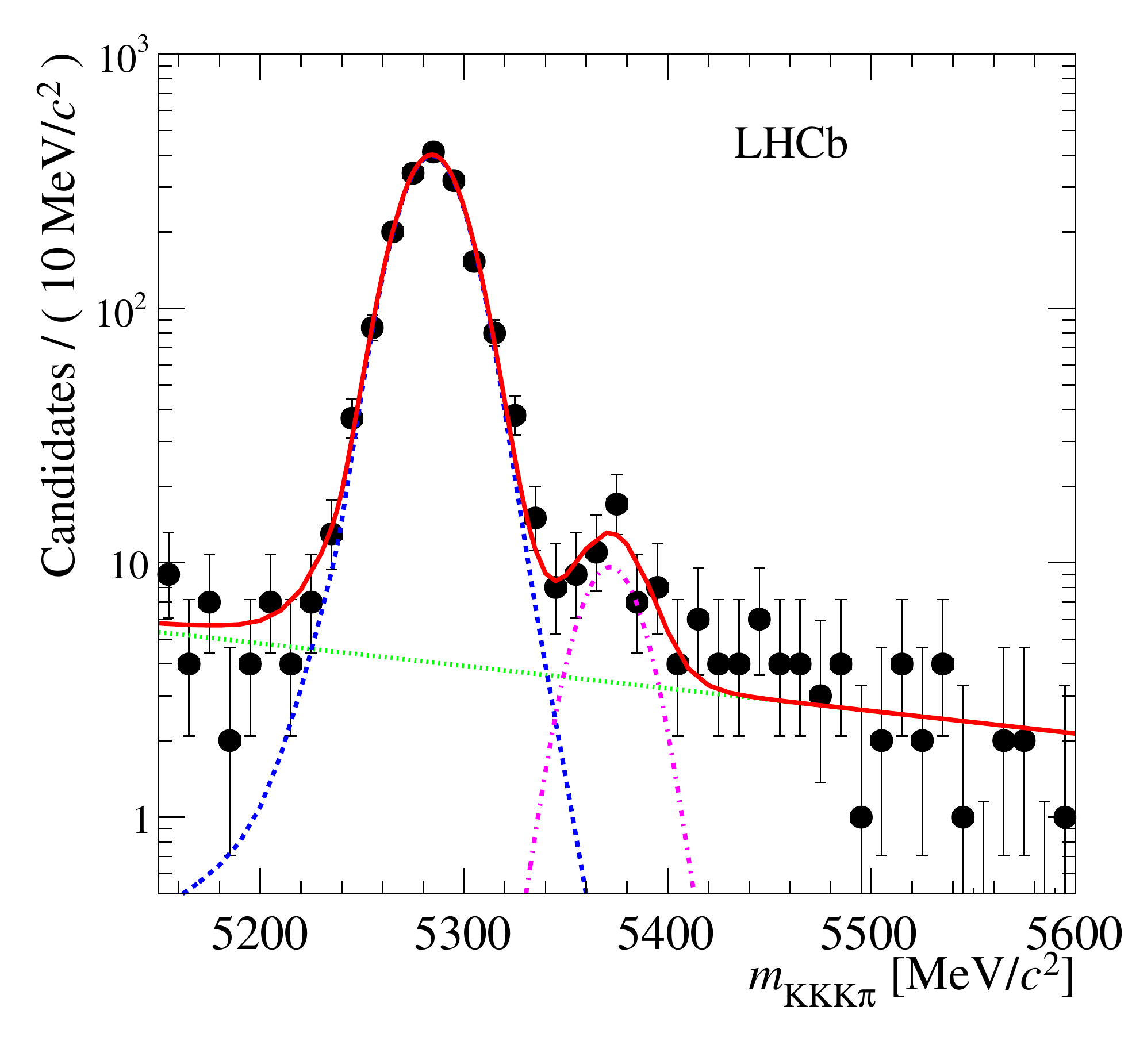}
\caption{\small Invariant mass distribution for selected $\KKKpi$
  candidates. A fit to the model described in the text is superimposed 
  (red solid line). The signal contribution is shown as the blue
  dotted line. The contribution from combinatorial background is shown in green 
  (dotted line). A contribution from \BsPhiKstb (purple dot-dashed line) decays 
  is visible around the known $B_s^0$ meson mass.}
\label{fig:bdmass}
\end{center}
\end{figure}

The signal yield is determined by an unbinned maximum likelihood fit to the $\KKKpi$ invariant mass distribution.
The selected mass range is chosen to avoid modelling partially reconstructed $B$ decays with a missing hadron or photon.
In the fit the signal invariant mass distribution is modelled as the sum of
a Crystal Ball function \cite{Skwarnicki:1986xj} and a wider Gaussian function with a common mean. 
The width and fraction of the Gaussian function are fixed to values obtained using
simulated events. A component is also included to account for
the small contribution from the decay \BsPhiKstb 
\cite{LHCb-PAPER-2013-012}. The shape parameters for this
component are in common with the $\Bz$ signal shape and the relative position of the $\Bs$ signal 
with respect to the $\Bz$ signal is fixed using the known mass difference between \Bz and \Bs mesons \cite{PDG2012}. The invariant mass distribution is shown in Fig.~\ref{fig:bdmass},
together with the result of the fit, from which a yield of $1655 \pm
42$ $B^0$ signal candidates is found.

After the selection the background is mainly combinatorial and is modelled by an exponential. Background from
$\Bs\rightarrow \phi \phi$ decays, with one of the kaons misidentified
as a pion is reduced  by the veto applied in the selection.  The number of candidates
from this source is estimated to be 6 events using simulation. These are distributed across 
the \KKKpi mass range, and are considered negligible in the fit. 
A potential background from $\Bz\rightarrow D_s^+ K^- (D_s^+\to \phi\pi^+)$ decays, 
which would peak in the signal region, is also found to be negligible. Possible background from the yet unobserved decay 
$\Lambda_b^0 \rightarrow \phi \proton K^-$ with a misidentified proton is considered as part of the systematic uncertainties. 

%% file: fitmodel.tex
\section{Angular fit}
\label{sec:fitmodel}
The physics parameters of interest for this analysis are defined in Table~\ref{tab:parameters}. They include the polarization amplitudes, 
phases and amplitude differences between \Bz and
\Bzb decays from which the triple-product asymmetries are calculated.

The correlation between the fit variables and $m_{KKK\pi}$ is found to
be less than $3 \, \%$. Therefore, the background can be subtracted
using the \sPlot\ method~\cite{Pivk:2004ty}, with $m_{\KKKpi}$ as the discriminating variable. The results of the invariant mass fit discussed in Sec.~\ref{sec:massmodel}
are used to give each candidate a signal weight, $W_n$, which is a
function of $m_{\KKKpi}$. The weight is used to subtract the background contributions 
from the distributions of the decay angles and intermediate resonance masses, which are fit using a signal-only likelihood that is a function of $\theta_1, \theta_2,
\Phi, m_{K\pi}$ and $m_{KK}$.
The angular fit minimizes the negative log likelihood summed over the $n$ selected candidates 
\begin{equation}
- \textrm{ln} \mathcal{L} = -\alpha \sum_{n} W_n \textrm{ln} \mathcal{S}_n\;,
\end{equation} 
where $\alpha = \sum_{n} W_{n}/\sum_{n} W_n^2$ is a normalization
factor that includes the effect of the weights in the determination of
the uncertainties \cite{LHCb-PAPER-2012-032,Eadie-71}, and
$\mathcal{S}$ is the signal probability density function
(Eq.~\ref{eqn:decay_sum}) convolved with the detector acceptance. 

The acceptance of the detector is not uniform as a function of
the decay angle of the $K^+\pi^-$ system~($\theta_1$) and 
the $K^+\pi^-$ invariant mass. This is due to the $500 \, \mevc$ criterion applied on the 
$\pt$ of the pion from the \Kstarz meson decay. In contrast, the acceptance is relatively
uniform as a function of the decay angles $\theta_2$ and $\Phi$, and the invariant mass of the $K^+K^-$ system.

\begin{figure}
\begin{center}
\includegraphics[width= 0.32\columnwidth]{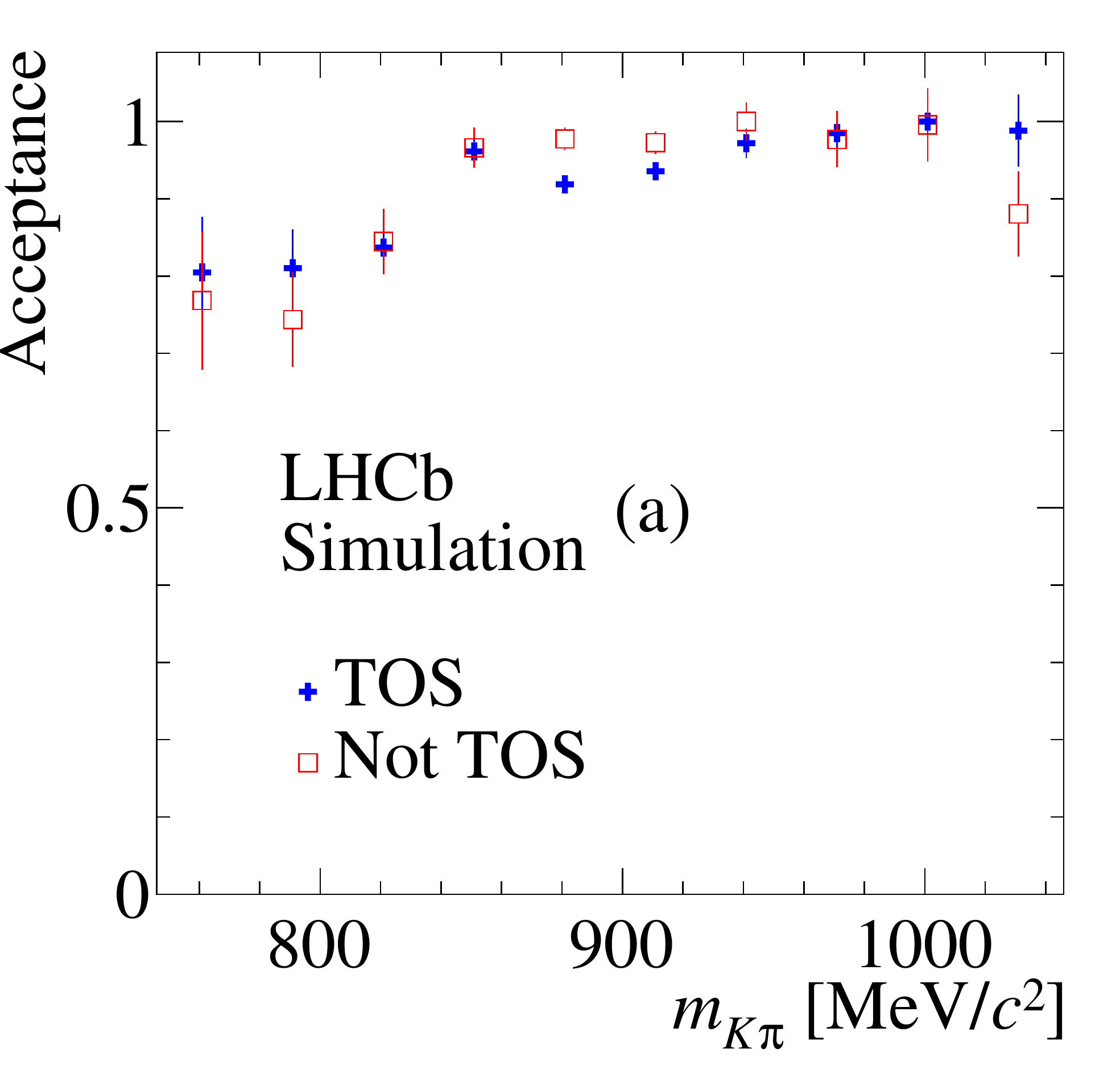}
\includegraphics[width= 0.32\columnwidth]{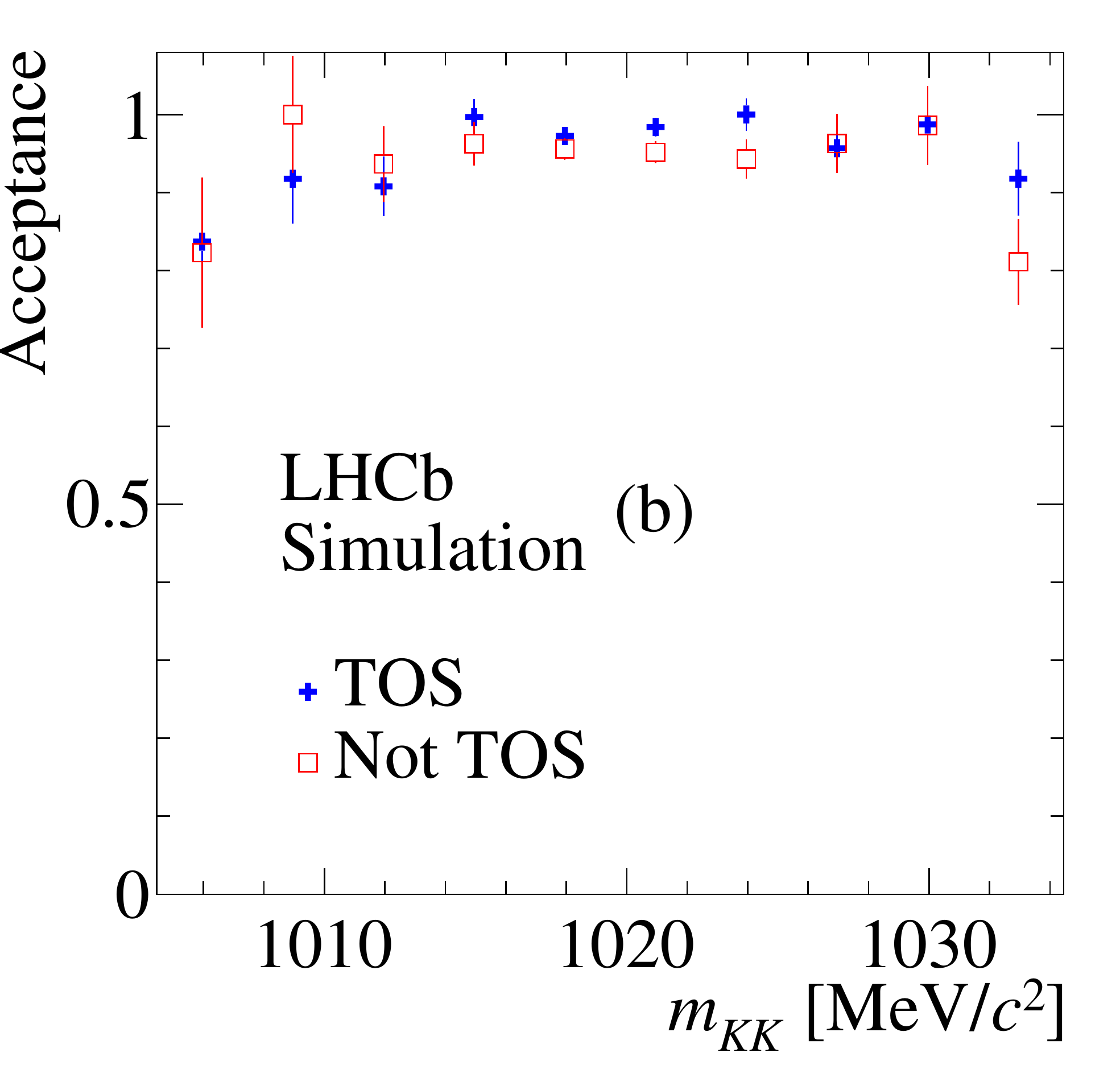}
\includegraphics[width= 0.32\columnwidth]{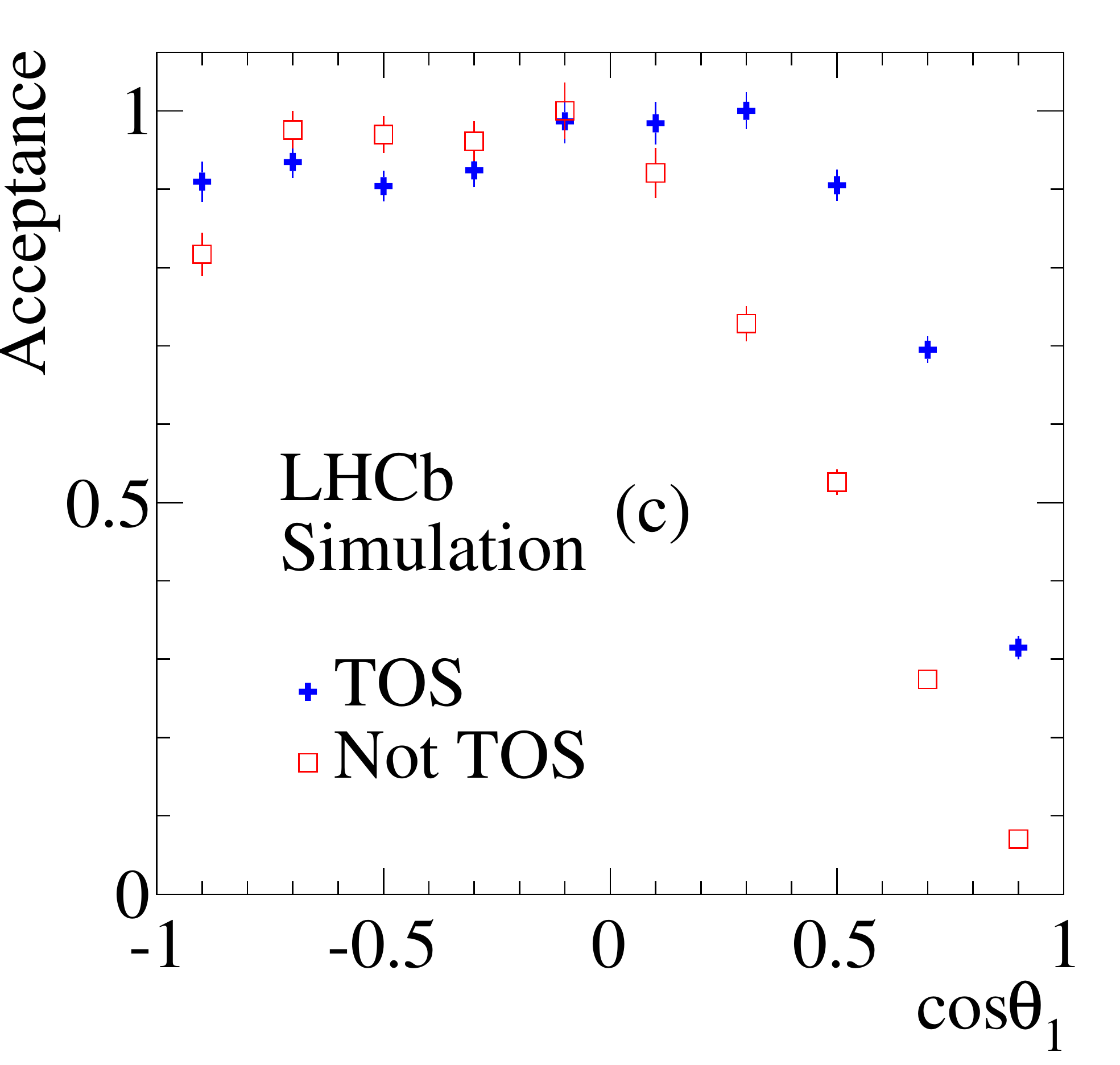}
\includegraphics[width= 0.32\columnwidth]{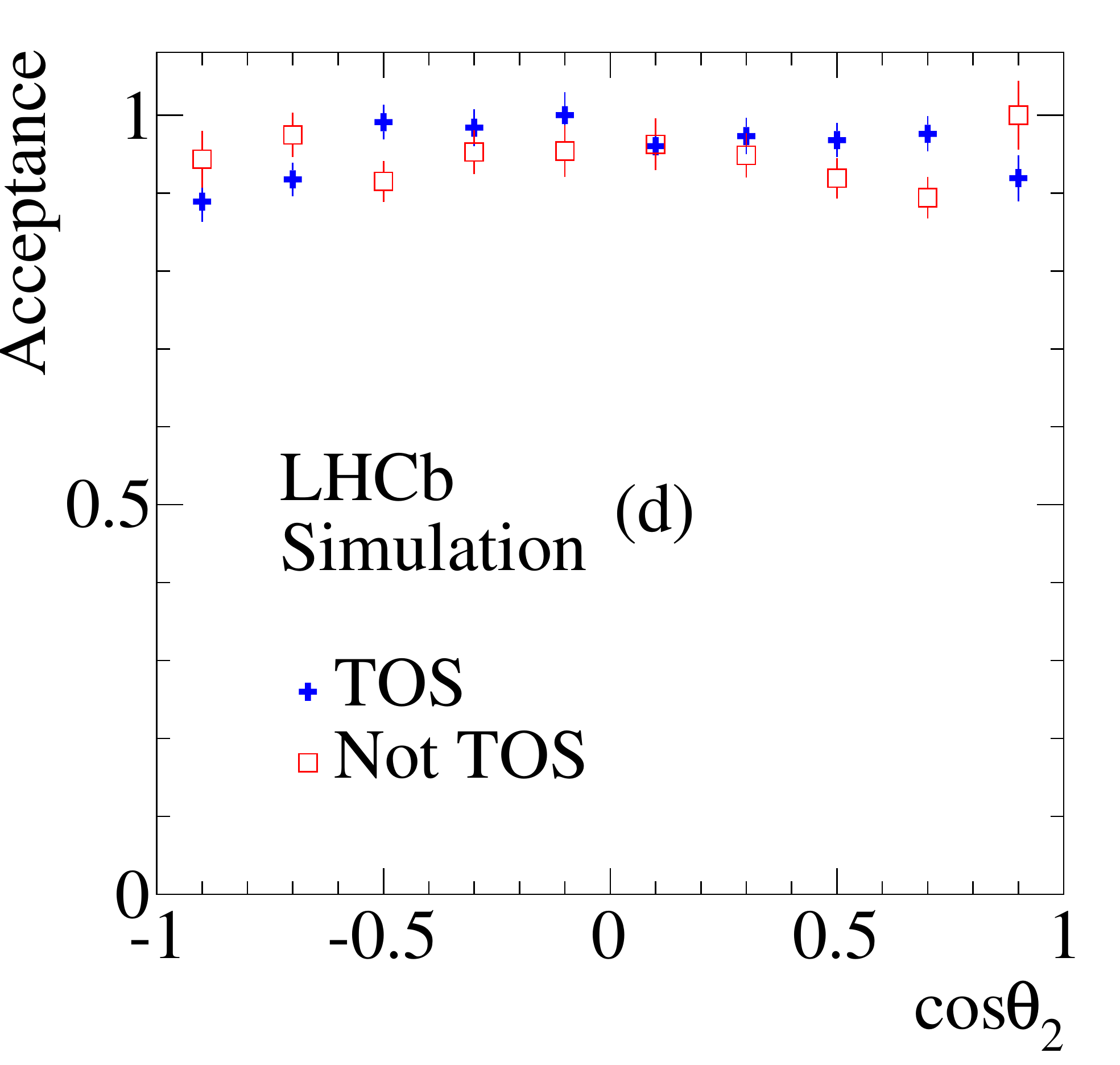}
\includegraphics[width= 0.32\columnwidth]{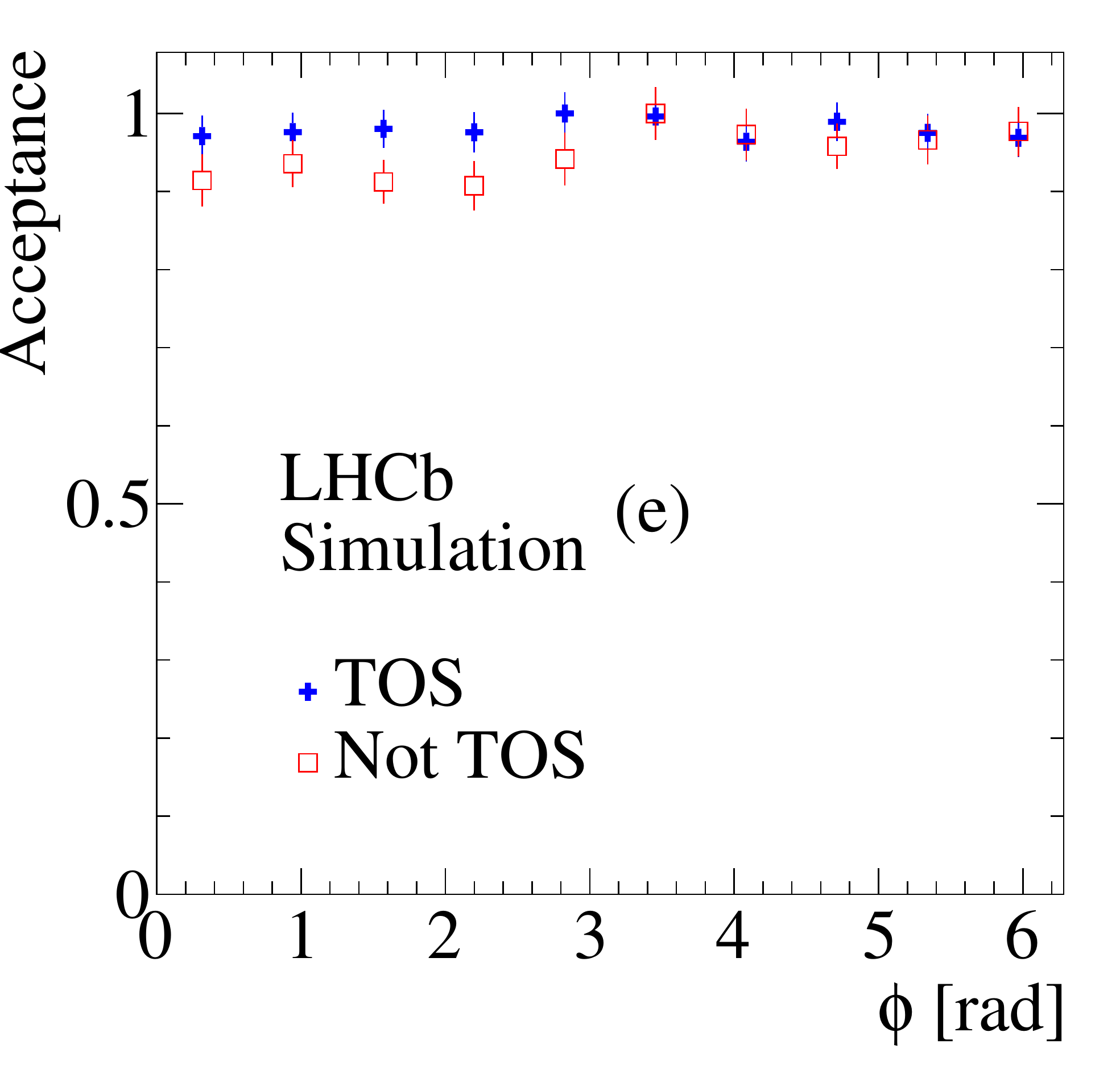}
\caption{\small Binned projections of the detector acceptance for (a) $m_{K\pi}$, (b) $m_{KK}$ , (c)
  $\cos\theta_1$, (d) $\cos\theta_2$ and (e) $\Phi$. The acceptance for the TOS~(filled crosses) and not TOS~(open squares) are shown on each plot.
\label{fig:acc_pk}}
\end{center}
\end{figure}

The detector acceptance is modelled using a four-dimensional function that depends on the 
three decay angles and the $K^+\pi^-$ invariant mass. 
The shape of this function is obtained from simulated data. 
As the quantities relating to the $\pt$ of the decay products are used in the
first-level hardware based trigger, the acceptance is different for
candidates that have a TIS or TOS decision at the hardware trigger
stage \cite{LHCb-DP-2012-004}. Consequently, the trigger acceptance is calculated and
corrected separately for the two categories. The $17 \, \%$ of candidates that fall in the
overlap between the two categories are treated as TOS, and the remaining TIS candidates are
labelled `not TOS'. The projections of the acceptance are shown in Fig.~\ref{fig:acc_pk}. 
In the subsequent analysis the data set is divided into the two categories and a 
simultaneous fit is performed. 

%% file: results.tex
\section{Angular analysis results}
\label{sec:results}  
\begin{figure}[htb!]
\begin{center}
\includegraphics[width = 0.48\columnwidth]{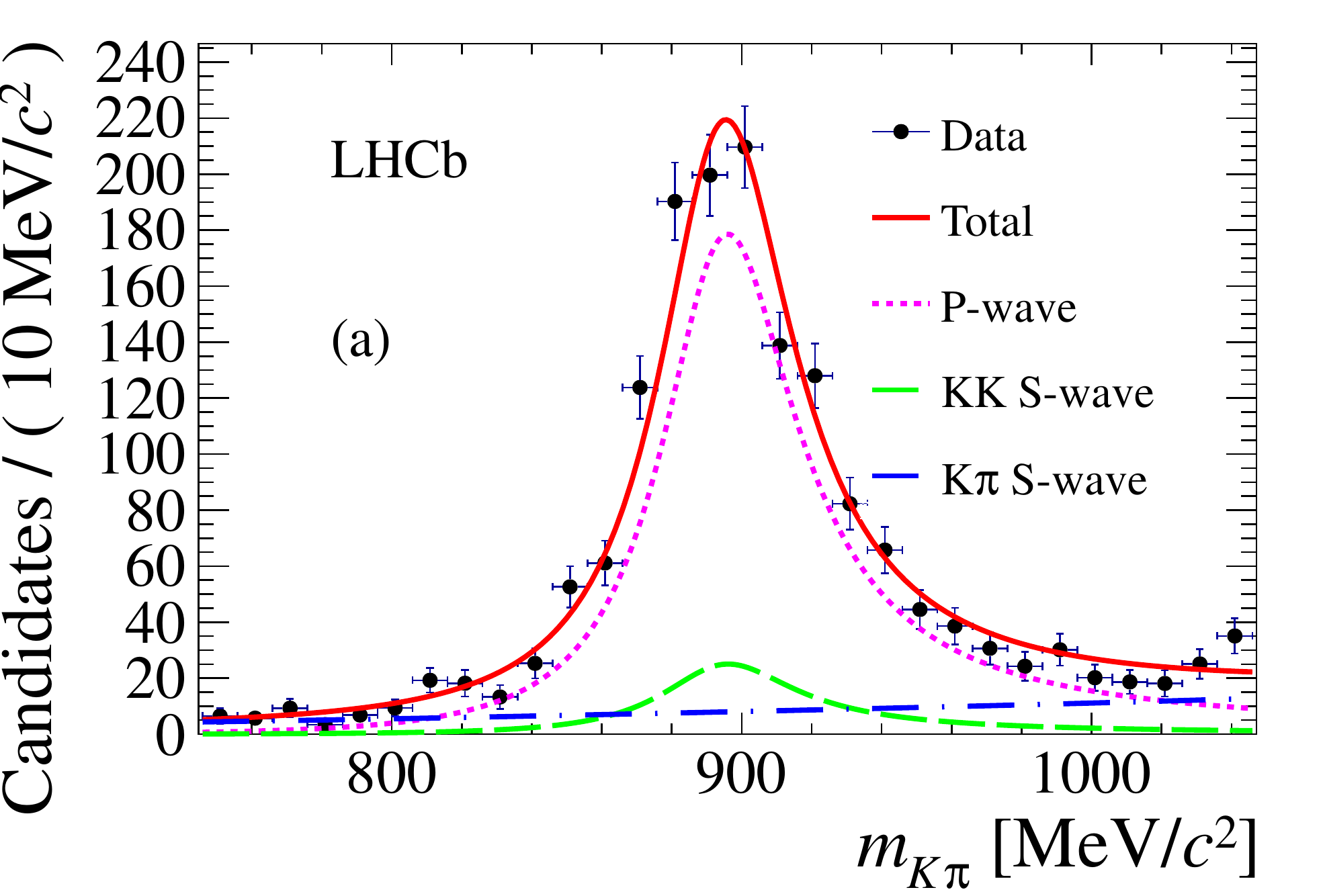}
\includegraphics[width = 0.48\columnwidth]{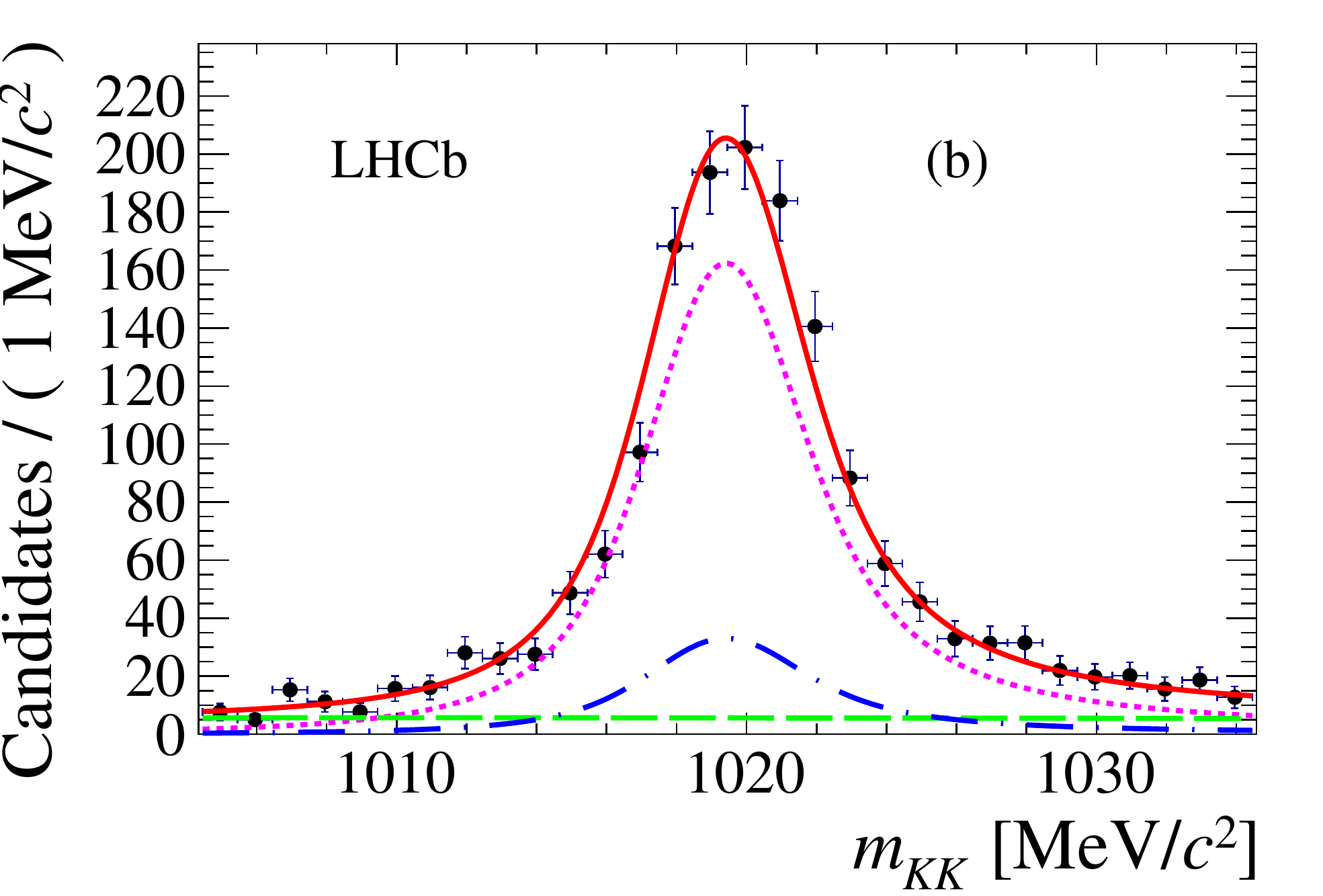}
\includegraphics[width = 0.48\columnwidth]{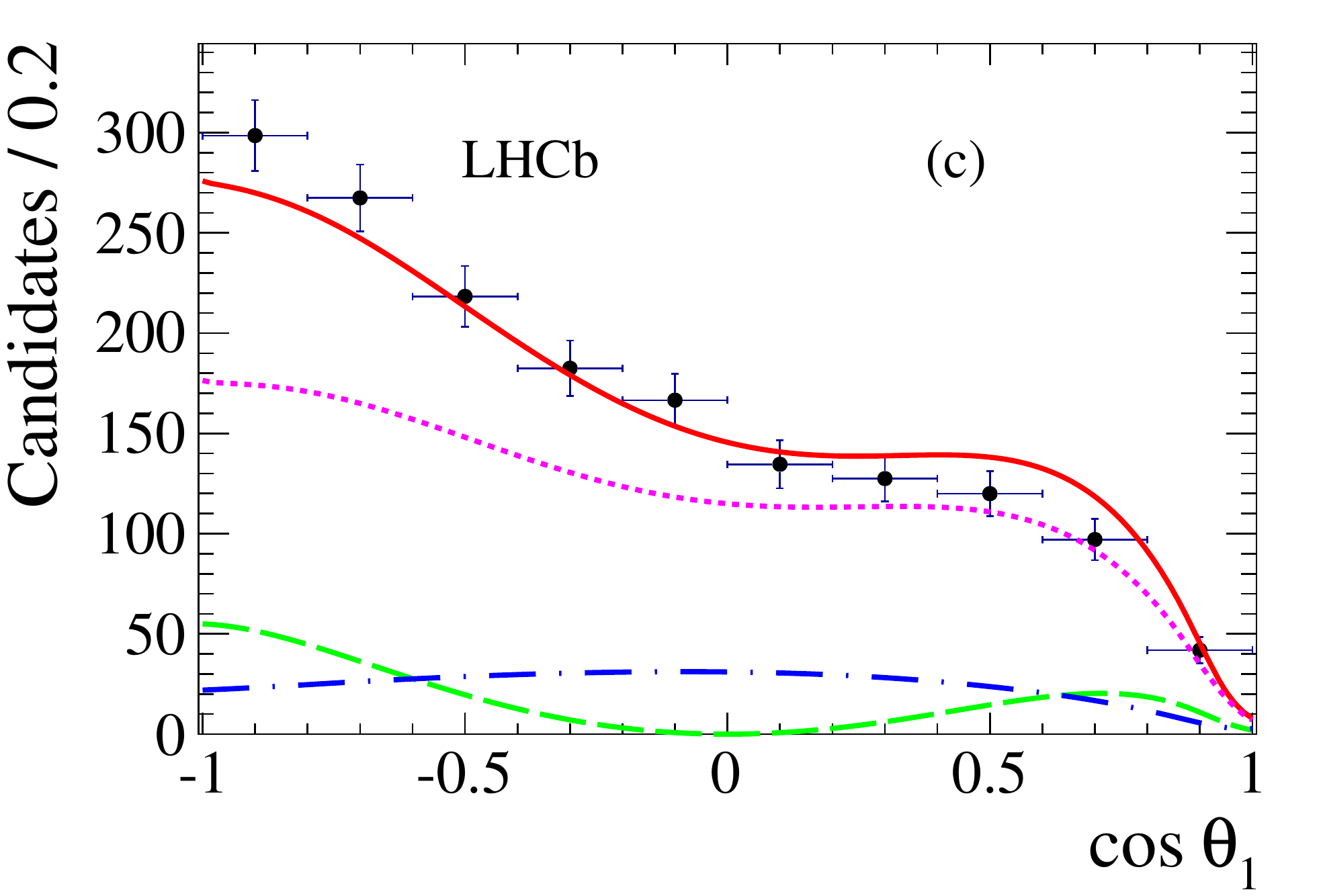}
\includegraphics[width = 0.48\columnwidth]{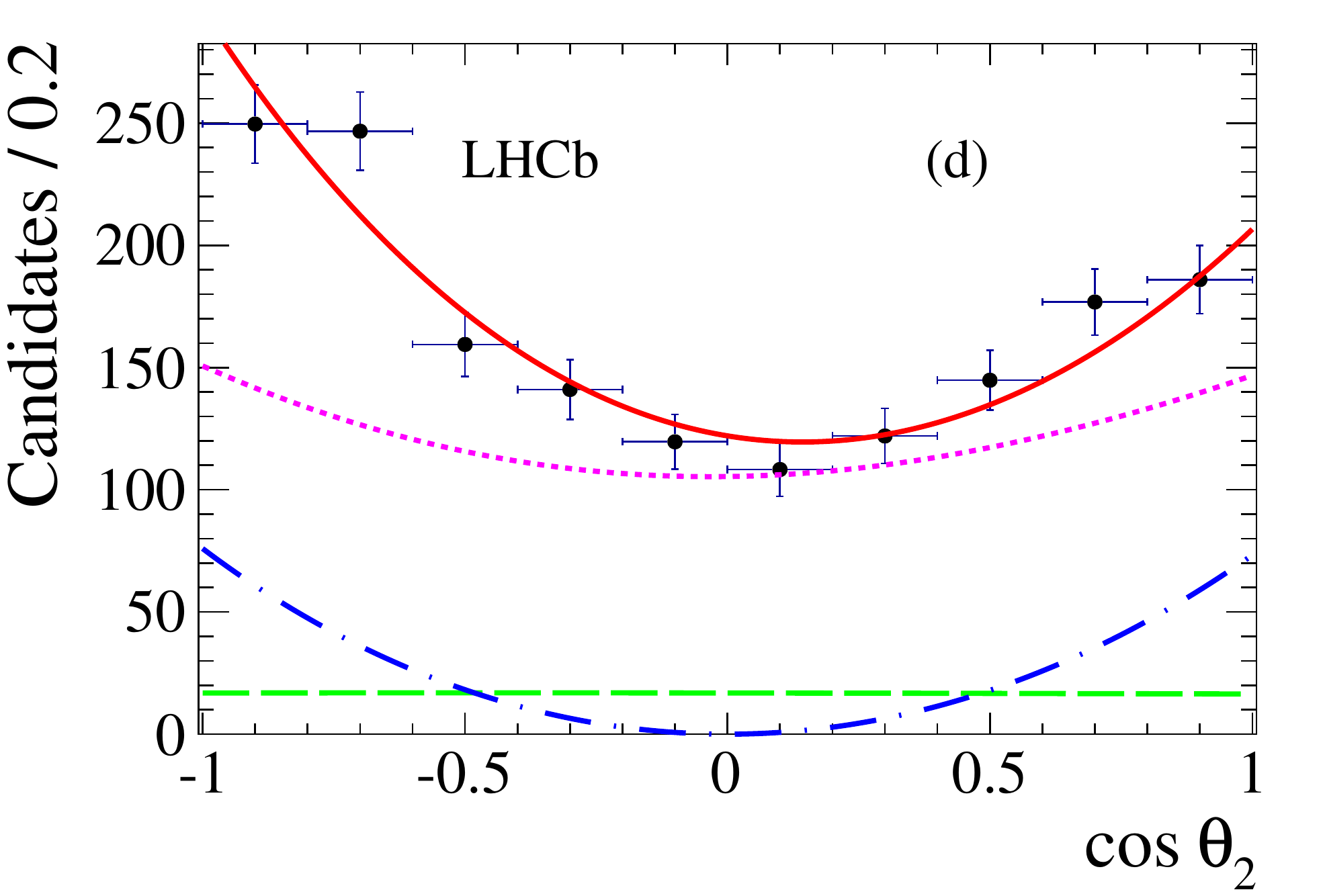}
\includegraphics[width = 0.48\columnwidth]{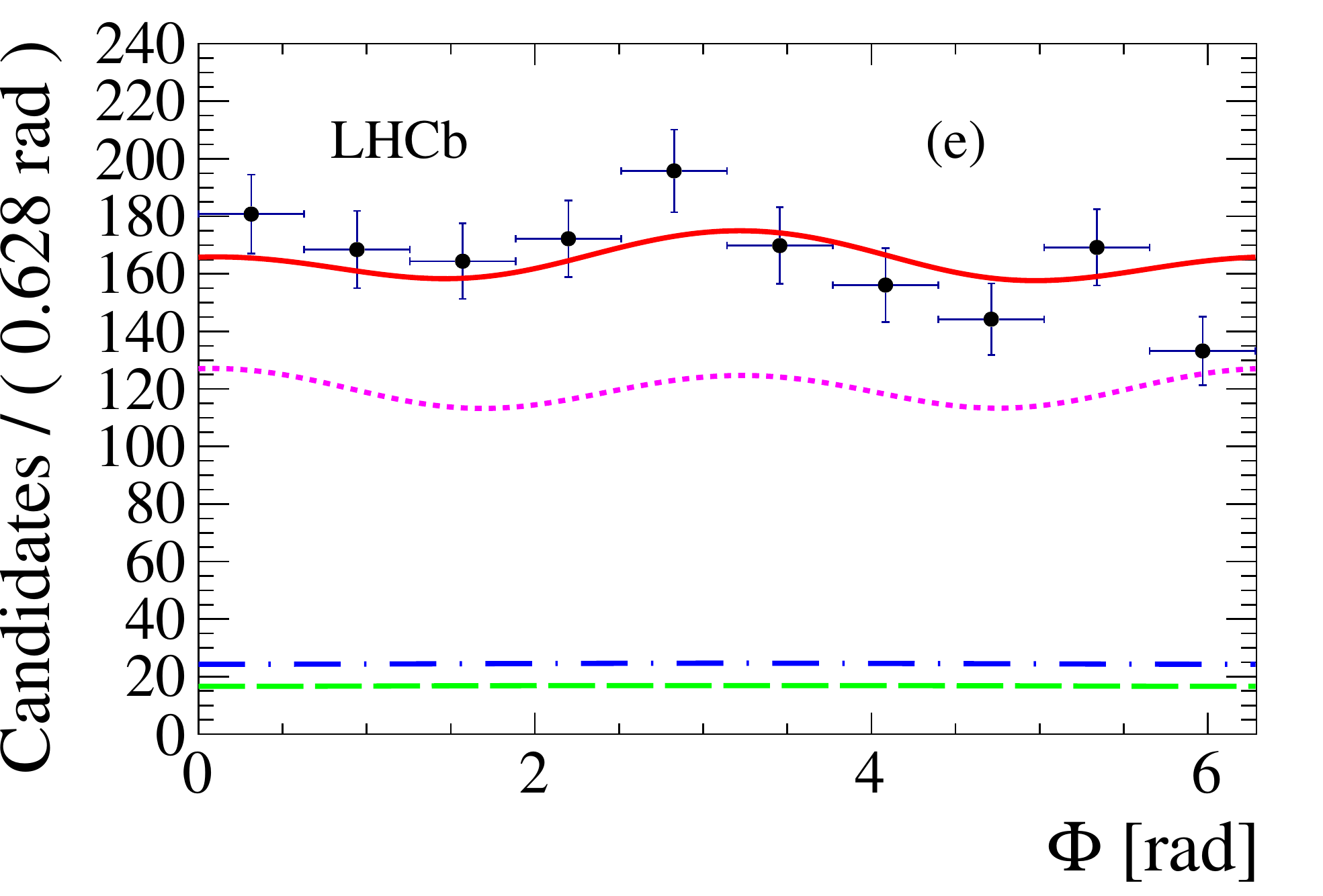}
\caption{\small Data distribution for the helicity angles and of the
intermediate resonance masses:  (a) $m_{K\pi}$ and (b) $m_{KK}$, (c) $\cos \theta_1$, (d) $\cos\theta_2$ and 
(e) $\Phi$. The background has been 
subtracted using the \sPlot\ technique. The results of the fit are superimposed.
\label{fig:fitproj}}
\end{center}
\end{figure}

Figure \ref{fig:fitproj} shows the data distribution for the intermediate resonance masses and helicity angles with the projections of the best fit
overlaid. The goodness of fit is estimated using a point-to-point dissimilarity
test \cite{Williams:2010vh}, the corresponding $p$-value is 0.64. 

The fit results are listed in Table~\ref{tab:parameters}. The value of
$f_L$  returned by the fit is close to 0.5, indicating that the
longitudinal and transverse polarizations have similar size. Significant S-wave 
contributions are found in both the $K^+\pi^-$ and $K^+K^-$
systems. The \CP asymmetries in both the amplitudes and 
the phases are consistent with zero. 

Using Eqs. \ref{for:TP12} -- \ref{for:TP34}, the  values for the
triple-product asymmetries are derived from the measured parameters 
and given in Table~\ref{tab:tripleproducts}. 
The true asymmetries are consistent with zero, showing no evidence for physics beyond the Standard Model. 
In contrast, all but one of the fake asymmetries are significantly different from zero, 
indicating the presence of final-state interactions. 

The systematic uncertainties on the measured amplitudes, phases and triple-product asymmetries are summarized in Table~\ref{tab:systematics}.
The largest systematic uncertainties on the results of the angular analysis 
arise from the understanding of the detector acceptance. 
The angular acceptance function is determined from simulated
events as described in Sec.~{\ref{sec:fitmodel}}. An uncertainty,
labelled `Acceptance' in the table, is assigned to account for the limited size of the simulation
sample used. This is estimated using pseudo-experiments
with a simplified simulation.

A difference is observed in the kinematic distributions
of the final-state particles between data and simulation. 
This is attributed to the S-wave components, which are
not included in the simulation. To account for this,
the simulated events are reweighted to match the signal distributions as 
expected from the best estimate of the physics parameters from data
(including the S-wave). In addition, the events are reweighted to match the 
observed distributions of the \Bz candidate and final-state particle transverse momenta. 
The reweighting is done separately for the two trigger categories and
the nominal results are recalculated using the reweighted simulation to determine the angular
acceptance. The difference 
between the weighted and unweighted results is taken as a systematic
uncertainty (labelled `Data/MC' in the table). 

A further uncertainty arises from the $\KKKpi$ mass model used to
determine the signal weights for the angular analysis. The fit procedure is repeated using
different signal and background models. For the
signal component a double Gaussian model is used instead of the sum of 
a Gaussian and a Crystal Ball function. Similarly, the influence of
background modelling is probed using a first-order polynomial
instead of an exponential function. Other changes to the background
model are related to the possible presence of additional backgrounds. 
A possible small contribution from misidentified $\Lambda_b
\to pK^-K^+K^-$ and $\Lambda_b \to p \pi^- K^+K^+$ decays is added and the
fit repeated. Finally, the lower bound of the fit range is
varied and the contribution from partially
reconstructed $B$ decays modelled. The largest difference 
compared to the central values is assigned as an estimate of the systematic
uncertainty (labelled `Mass model' in the table). 

Alternative models of the S-wave contributions in both the $K^+K^-$
and $K^+\pi^-$ system are considered. The default fit uses the LASS parameterization to model the $K^+\pi^-$
S-wave. As variations of this, both a pure phase-space model
and a spin-0 relativistic Breit-Wigner with mean and width of the
$K^*_0(1430)$ resonance are considered~{\cite{PDG2012}}. For the $K^+K^-$ S-wave
a pure phase-space model is tried in place of the Flatt\'e
parameterization. The largest observed deviation from the
nominal fit is taken as a systematic uncertainty (column  labelled
`S-wave' in the table).

Various consistency checks of the results are made. As a cross-check
candidates that are in the overlap between the trigger categories are
treated as TIS for the angular correction in the fit rather than TOS. The dataset is also divided according to the
magnetic field polarity. The results obtained in these studies are
consistent with the nominal results and no additional uncertainty is assigned.

\begin{table}[!htb]
\begin{center}
\caption{\small Parameters measured in the angular analysis. 
The first and second uncertainties are statistical and systematic, respectively. 
\label{tab:parameters}}
\begin{tabular}{ccc}
\hline
Parameter & Definition & Fitted value \\ 
\hline 
\\ [-0.48cm]
$f_\textrm{L}$ & $0.5(|A_0|^2/F_\textrm{P} + |\overline{A}_0|^2/\overline{F}_\textrm{P})$ &\phantom{-} 
$0.497\pm 0.019\pm 0.015$ \\
$f_{\perp}$ & $0.5(|A_\perp|^2/F_\textrm{P} + |\overline{A}_\perp|^2/\overline{F}_\textrm{P})$ & \phantom{-}
$0.221\pm 0.016\pm 0.013$ \\
$f_{\textrm{S}}(K\pi)$  & $0.5(|A_{\textrm{S}}^{K\pi}|^2 + |\overline{A}_{\textrm{S}}^{K\pi}|^2)$ & \phantom{-}
$0.143\pm 0.013\pm 0.012$\\
$f_{\textrm{S}}(KK)$  & $0.5(|A_{\textrm{S}}^{KK}|^2 + |\overline{A}_{\textrm{S}}^{KK}|^2)$ & \phantom{-}
$0.122\pm 0.013\pm 0.008$ \\
\\ [-0.48cm]
$\delta_\perp$     & $0.5(\arg{A_\perp} + \arg{\overline{A}_\perp})$ & \phantom{-}
$2.633\pm 0.062\pm 0.037$ \\
$\delta_\parallel$     & $0.5(\arg{A_\parallel} + \arg{\overline{A}_\parallel})$ &\phantom{-} 
$2.562\pm 0.069\pm 0.040$ \\
$\delta_{\textrm{S}}(K\pi)$     & $0.5(\arg{A_{\textrm{S}}^{K\pi}} + \arg{\overline{A}_{\textrm{S}}^{K\pi}})$ & \phantom{-}
$2.222\pm 0.063\pm 0.081$ \\
$\delta_{\textrm{S}}(KK)$     & $0.5(\arg{A_{\textrm{S}}^{KK}} + \arg{\overline{A}_{\textrm{S}}^{KK}})$ & \phantom{-}
$2.481\pm 0.072\pm 0.048$\\ 
\\ [-0.48cm]
$\mathcal{A}_{0}^{\CP}$         & $(|A_0|^2{/F_\textrm{P}} -
|\overline{A}_0|^2{/\overline{F}_\textrm{P}})/(|A_0|^2{/F_\textrm{P}}
+ |\overline{A}_0|^2{/\overline{F}_\textrm{P}})$ & 
$-0.003\pm 0.038\pm 0.005$ \\
$\mathcal{A}_{\perp}^{\CP}$         & $(|A_\perp|^2{/F_\textrm{P}} -
|\overline{A}_\perp|^2{/\overline{F}_\textrm{P}})/(|A_\perp|^2{/F_\textrm{P}}
+ |\overline{A}_\perp|^2{/\overline{F}_\textrm{P}})$ &
$+0.047\pm 0.074\pm 0.009$ \\
$\mathcal{A}_{S}(K\pi)^{\CP}$ & $(|A_{\textrm{S}}^{K\pi}|^2 - |\overline{A}_{\textrm{S}}^{K\pi}|^2)/(|A_{\textrm{S}}^{K\pi}|^2 + |\overline{A}_{\textrm{S}}^{K\pi}|^2)$ &
$+0.073\pm 0.091\pm 0.035$ \\
$\mathcal{A}_{S}(KK)^{\CP}$   & $(|A_{\textrm{S}}^{KK}|^2 - |\overline{A}_{\textrm{S}}^{KK}|^2)/(|A_{\textrm{S}}^{KK}|^2 + |\overline{A}_{\textrm{S}}^{KK}|^2)$ & 
$-0.209\pm 0.105\pm 0.012$ \\
\\ [-0.48cm]
$\delta_\perp^{\CP}$          & $0.5(\arg{A_\perp} - \arg{\overline{A}_\perp})$ & 
$+0.062\pm 0.062\pm 0.005$ \\
$\delta_\parallel^{\CP}$      & $0.5(\arg{A_\parallel} - \arg{\overline{A}_\parallel})$ & 
$+0.045\pm 0.069\pm 0.015$ \\
$\delta_{S}(K\pi)^{\CP}$      & $0.5(\arg{A_{\textrm{S}}^{K\pi}} - \arg{\overline{A}_{\textrm{S}}^{K\pi}})$ & 
$+0.062\pm 0.062\pm 0.022$ \\
$\delta_{S}(KK)^{\CP}$        & $0.5(\arg{A_{\textrm{S}}^{KK}} - \arg{\overline{A}_{\textrm{S}}^{KK}})$ &
$+0.022\pm 0.072\pm 0.004$ \\\hline 
\end{tabular}
\end{center}
\end{table}

\begin{table}[!htb]
\begin{center}
\caption{\small Triple-product asymmetries. The first and second uncertainties on the measured values 
are statistical and systematic, respectively. \label{tab:tripleproducts}}
\begin{tabular}{cc}
\hline
Asymmetry & Measured value \\ 
\hline 
$A_{T}^1$(true) & $-0.007\pm 0.012\pm 0.002$ \\
$A_{T}^2$(true) & $+0.004\pm 0.014\pm 0.002$ \\
$A_{T}^3$(true) & $+0.004\pm 0.006\pm 0.001$ \\
$A_{T}^4$(true) & $+0.002\pm 0.006\pm 0.001$ \\ 
$A_{T}^1$(fake) & $-0.105\pm 0.012\pm 0.006$ \\
$A_{T}^2$(fake) & $-0.017\pm 0.014\pm 0.003$ \\
$A_{T}^3$(fake) & $-0.063\pm 0.006\pm 0.005$ \\
$A_{T}^4$(fake) & $-0.019\pm 0.006\pm 0.007$ \\  \hline
\end{tabular}
\end{center}
\end{table}

\begin{table}[ht]
\caption{\small Systematic uncertainties on the measurement of
the polarization amplitudes, relative strong phases and triple-product
asymmetries. The column labelled `Total' is the quadratic sum of the
individual contributions. \label{tab:systematics}}
\begin{center}
\begin{tabular}{lccccc} 
\hline
Measurement                        & Acceptance & Data/MC  & Mass model & S-wave &  Total \\
\hline                                                    
$f_\textrm{L}$                     & 0.014       & 0.005   & 0.002      & 0.001  & 0.015\\
$f_\perp$                          & 0.013       & 0.002   & 0.001      & 0.001  & 0.013\\
$f_{\textrm{S}}(K\pi)$             & 0.012       &   -     & 0.001      & 0.002  & 0.012\\
$f_{\textrm{S}}(KK)$               & 0.007       &   -     & 0.002      & 0.003  & 0.008\\
$\delta_\perp$                     & 0.023       & 0.010   & 0.006      & 0.026  & 0.037\\
$\delta_\parallel$                 & 0.029       & 0.013   & 0.004      & 0.024  & 0.040\\
$\delta_{\textrm{S}}(K\pi)$        & 0.045       & 0.026   & 0.004      & 0.062  & 0.081\\
$\delta_{\textrm{S}}(KK)$          & 0.045       & 0.005   & 0.004      & 0.016  & 0.048\\
$A_0^{\CP}$                        & -           & 0.002   & 0.002      & 0.004  & 0.005\\
$A_\perp^{\CP}$                    & -           & 0.001   & 0.006      & 0.007  & 0.009\\
$A_{\textrm{S}}(K\pi)^{\CP}$       & -           & 0.007   & 0.005      & 0.034  & 0.035\\
$A_{\textrm{S}}(KK)^{\CP}$         & -           & 0.007   & 0.009      & 0.003  & 0.012\\
$\delta_\perp^{\CP}$               & -           & 0.003   & 0.001      & 0.004  & 0.005\\
$\delta_\parallel^{\CP}$           & -           & 0.005   & 0.002      & 0.014  & 0.015\\
$\delta_{\textrm{S}}(K\pi)^{\CP}$  & -           & 0.005   & 0.003      & 0.021  & 0.022\\
$\delta_{\textrm{S}}(KK)^{\CP}$    & -           & 0.002   & 0.002      & 0.003  & 0.004\\
$A_{T}^1$(true)                    & -           & 0.0005  & 0.0005     & 0.002  & 0.002\\
$A_{T}^2$(true)                    & -           & 0.0006  & 0.0005     & 0.002  & 0.002\\
$A_{T}^3$(true)                    & -           & 0.0002  & 0.0003     & 0.001  & 0.001\\
$A_{T}^4$(true)                    & -           & 0.0002  & 0.0003     & 0.001  & 0.001\\
$A_{T}^1$(fake)                    & -           & 0.0019  & 0.0017     & 0.005  & 0.006\\
$A_{T}^2$(fake)                    & -           & 0.0008  & 0.0008     & 0.003  & 0.003\\
$A_{T}^3$(fake)                    & -           & 0.0015  & 0.0006     & 0.005  & 0.005\\
$A_{T}^4$(fake)                    & -           & 0.0003  & 0.0004     & 0.007  & 0.007\\ \hline 
\end{tabular}
\end{center}
\end{table}

%% file: acp.tex
\section{Direct \boldmath \CP rate asymmetry}
\label{sec:acp}
The raw measurement of the rate asymmetry is obtained from
\begin{equation}
A = \frac{N(\BdPhiKstb)-N(\BdPhiKst)}
{N(\BdPhiKstb)+N(\BdPhiKst)}\;.
\end{equation}
The numbers of events, $N$, are determined from fits to the $m_{KKK\pi}$ invariant mass distribution 
performed separately for \Bz and \Bzb decays, identified using the charge of
the final-state kaon. The dilution from the S-wave components is
corrected for using the results of the angular analysis.

The candidates are separated into the TIS and TOS trigger
categories. In this study, candidates that are
accepted by both trigger decisions are included in both categories and
a possible bias to the central value is treated as a systematic
uncertainty. The obtained raw asymmetries for the two trigger types are
\begin{equation}
A^{\textrm{TOS}}_{\phi K^{*0}} = +0.014\pm 0.043 \hspace{0.5cm} \textrm{and} \hspace{0.5cm} A^{\textrm{TIS}}_{\phi K^{*0}} = -0.002\pm 0.040\;. \nonumber
\end{equation}
The direct \CP asymmetry is related to the measured $A$ by 
\begin{equation}
A_{\CP} = A - \delta \hspace{0.5cm} \textrm{with} \hspace{0.5cm} \delta = A_\textrm{D} +  \kappa_d A_\textrm{P}\;,
\end{equation}
\noindent where $A_\textrm{D}$ is the detection asymmetry between $K^+\pi^-$ and $K^-\pi^+$ final-states, 
$A_\textrm{P}$ is the asymmetry in production rate between \Bz and \Bzb mesons in $pp$ collisions, 
and the factor $\kappa_d$ accounts for the dilution of the production
asymmetry due to $\Bz-\Bzb$ oscillations.

The decay $\Bz \rightarrow J/\psi K^{*0}$ is used as a control channel to determine the difference in asymmetries
\begin{equation}
\Delta A_{\CP} = A_{\CP}(\phi K^{*0}) - A_{\CP}(J/\psi K^{*0})\;,
\end{equation}
since the detector and production asymmetries cancel in the
difference. Assuming $A_{CP}$ to be zero for the tree-level $B^0 \to J/\psi K^{*0}$ decay,
$\Delta A_{\CP}$ is the \CP asymmetry in \BdPhiKst.
The sample of $\Bz \rightarrow J/\psi K^{*0}$ decays, where the
$J/\psi$ meson decays to a muon pair, are collected through the same trigger and offline selections used for the signal decay mode. 
Candidates are placed in the TOS trigger category if the trigger decision is based on the decay products from the $K^{*0}$ meson only. Where the decay products from the $J/\psi$ meson influences the trigger decision, the candidate is rejected. The raw asymmetries obtained separately 
for the two trigger types are
\begin{equation}
A^{\textrm{TOS}}_{J/\psi K^{*0}} = -0.003\pm 0.016 \hspace{0.5cm} \textrm{and} \hspace{0.5cm} A^{\textrm{TIS}}_{J/\psi K^{*0}} = -0.016\pm 0.008\;. \nonumber
\end{equation}
After averaging the trigger categories based on their statistical uncertainty, the measured value for the difference in \CP asymmetries is
\begin{equation}
 \Delta A_{\CP} = (+1.5 \pm 3.2 \pm 0.5) \, \%\;, \nonumber
\end{equation} 
where the uncertainties are statistical and systematic, respectively. Systematic uncertainties arise from the
differences between the event topologies of the \mbox{$B^0 \rightarrow J/\psi
K^{*0}$} and \BdPhiKst decays. Differences in the behaviour of the
events in the TIS trigger category between the signal and control modes lead to an uncertainty of
$0.25 \, \%$. A further uncertainty of $0.4 \, \%$ arises from
the differences in kinematics of the daughter particles in the two modes. 
The double counting of candidates in the overlap region leads to a
possible bias on the central value, estimated to be less than
$0.1 \, \%$.

%% file: conclusions.tex
\section{Conclusions}
\label{sec:conclusions}
In this paper measurements of the polarization amplitudes and strong
phase differences in the decay mode \BdPhiKst are reported. The
results for the P-wave parameters are shown in Table~\ref{tab:compare}; these are consistent with, but more precise
than previous measurements. 
All measurements are consistent with the presence of a large transverse component 
rather than the na\"ive expectation of a dominant longitudinal polarization. 
\begin{table}[t]
\begin{center}
\caption{\small Comparison of measurements made by the \lhcb, \babar~\cite{Aubert:2008zza} and
  \belle~\cite{PhysRevD.88.072004} collaborations. The first uncertainty is statistical and the
  second systematic. \label{tab:compare}}
\begin{tabular}{lccc}
\hline
Parameter              & \lhcb           & \babar    & \belle  \\ \hline
$f_{\textrm{L}}$              & $\phantom{-}0.497\pm 0.019\pm 0.015$  &
$\phantom{-}0.494\pm 0.034\pm 0.013$  & $\phantom{-}0.499\pm 0.030\pm 0.018$ \\

$f_{\perp}$ & $\phantom{-}0.221\pm 0.016\pm 0.013$   &
$\phantom{-}0.212\pm 0.032\pm 0.013$ & $\phantom{-}0.238\pm 0.026\pm 0.008$  \\

$\delta_\perp$     & $\phantom{-}2.633\pm 0.062\pm 0.037$   &
$\phantom{-}2.35\phantom{0}\pm 0.13\phantom{0}\pm 0.09\phantom{0}$  & $\phantom{-}2.37\phantom{0}\pm 0.10\phantom{0}\pm 0.04\phantom{0}$ \\

$\delta_\parallel$      & $\phantom{-}2.562\pm 0.069\pm 0.040$   & 
$\phantom{-}2.40\phantom{0}\pm 0.13\phantom{0}\pm 0.08\phantom{0}$  & $\phantom{-}2.23\phantom{0}\pm 0.10\phantom{0}\pm 0.02\phantom{0}$  \\

$A_0^{\CP}$              & $-0.003\pm 0.038\pm 0.005$   & 
$+0.01\phantom{0}\pm 0.07\phantom{0}\pm 0.02\phantom{0}$  & $-0.030 \pm 0.061 \pm 0.007$ \\

$A_\perp^{\CP}$          & $+0.047\pm 0.072\pm 0.009$   & 
$-0.04\phantom{0}\pm 0.15\phantom{0} \pm 0.06\phantom{0}$   &  $-0.14\phantom{0} \pm 0.11\phantom{0} \pm 0.01\phantom{0}$ \\

$\delta_\perp^{\CP}$     & $+0.062\pm 0.062\pm 0.006$   &
$+0.21\phantom{0}\pm 0.13\phantom{0}\pm 0.08\phantom{0}$ & $+0.05\phantom{0} \pm 0.10\phantom{0} \pm 0.02\phantom{0}$ \\

$\delta_\parallel^{\CP}$ & $+0.045\pm 0.068\pm 0.015$   &
$+0.22\phantom{0}\pm 0.12\phantom{0}\pm 0.08\phantom{0}$   &  $-0.02\phantom{0} \pm 0.10\phantom{0} \pm 0.01\phantom{0}$ 
\\ \hline
\end{tabular}
\end{center}
\end{table}

It is more difficult to make comparisons for the S-wave components as this 
is the first measurement to include consistently the effect of the S-wave in the $K^+K^-$ system, and because 
the $K^+\pi^-$ mass range is different with respect to the range used in previous analyses. The measurements of the 
polarization amplitude differences are consistent with \CP conservation.  

The results of the angular analysis are used to determine triple-product asymmetries. The measured true asymmetries show no evidence for \CP
violation. In contrast, large fake asymmetries are observed, indicating the presence of significant final-state interactions. 
The difference in direct \CP asymmetries between the \BdPhiKst and $B^0 \to J/\psi K^{*0}$ decays is also measured,
\begin{eqnarray}
 \Delta A_{\CP} = (+1.5 \pm 3.2 \pm 0.5) \, \%\;, \nonumber
\end{eqnarray}
where the first uncertainty is statistical and the second systematic.
This is a factor of two more precise than previous values reported by \babar and \belle~{\cite{Aubert:2008zza,PhysRevD.88.072004}} and is found to be consistent with zero.

%% file: acknowledgements.tex
\section*{Acknowledgements}
\noindent We express our gratitude to our colleagues in the CERN
accelerator departments for the excellent performance of the LHC. We
thank the technical and administrative staff at the LHCb
institutes. We acknowledge support from CERN and from the national
agencies: CAPES, CNPq, FAPERJ and FINEP (Brazil); NSFC (China);
CNRS/IN2P3 and Region Auvergne (France); BMBF, DFG, HGF and MPG
(Germany); SFI (Ireland); INFN (Italy); FOM and NWO (The Netherlands);
SCSR (Poland); MEN/IFA (Romania); MinES, Rosatom, RFBR and NRC
``Kurchatov Institute'' (Russia); MinECo, XuntaGal and GENCAT (Spain);
SNSF and SER (Switzerland); NASU (Ukraine); STFC (United
Kingdom); NSF (USA). We also acknowledge the support received from EPLANET and the
ERC under FP7. The Tier1 computing centres are supported by IN2P3
(France), KIT and BMBF (Germany), INFN (Italy), NWO and SURF (The
Netherlands), PIC (Spain), GridPP (United Kingdom).
We are indebted to the communities behind the multiple open source software packages on which we depend.
We are also thankful for the computing resources and the access to software R\&D tools provided by Yandex LLC (Russia).